\documentclass[longauth,doublecolumn]{aa} 

\usepackage{graphicx}
\usepackage{txfonts}
\usepackage{savesym}
\savesymbol{rotate}

\usepackage[english]{babel}
\usepackage{times,epsfig}
\usepackage{longtable}
\usepackage{url}

\usepackage{mathrsfs}
\usepackage[normalem]{ulem}

\usepackage{soul}
\usepackage{ulem}

\usepackage{hyperref}
\hypersetup{
	colorlinks = true,
	linkcolor = blue,
	anchorcolor = blue,
	citecolor = blue,
	filecolor = blue,
	urlcolor = blue
}

\usepackage{placeins}
\usepackage{xcolor}

\newcommand{\orcid}[1]{\href{https://orcid.org/#1}{\includegraphics[width=10pt]{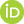}}}

\def\apjl{ApJL}%
\def\apjs{ApJS}%
\def\ao{Appl.~Opt.}%
\def\aap{A\&A}%

\DeclareMathAlphabet{\mathsc}{OT1}{cmr}{m}{sc}
\def\testbx{bx}%
\DeclareRobustCommand{\ion}[2]{%
\relax\ifmmode
\ifx\testbx\f@series
{\mathbf{#1\,\mathsc{#2}}}\else
{\mathrm{#1\,\mathsc{#2}}}\fi
\else\textup{#1\,{\mdseries\textsc{#2}}}%
\fi}

\newcommand{\dmb} {\mbox{$\Delta m_{15}(B)$}}
\newcommand{\sbv} {\mbox{$s_{BV}$}}

\newcommand{\Ci} {\ion{C}{i}}

\newcommand{\Oi} {\ion{O}{i}}

\newcommand{\Nai} {\ion{Na}{i}}

\newcommand{\Siii} {\ion{Si}{ii}}

\newcommand{\Caii} {\ion{Ca}{ii}}

\newcommand{\sn}{SN~2021hem}

\newcommand{\PeakEpoch}{JD~2459316.46}

\newcommand{\kpc}{\ensuremath{\mbox{~kpc}}}
\newcommand{\bv}{\mbox{$B\!-\!V$}}

\newcommand{\msun}{\mbox{M$_{\odot}$}}

\newcommand{\kms}{\mbox{$\rm{\,km\,s^{-1}}$}}

\newcommand{\nickel}{\mbox{$^{56}$Ni}}
\newcommand{\cobalt}{\mbox{$^{56}$Co}}
\newcommand{\iron}{\mbox{$^{56}$Fe}}

\newcommand{\mum}{\ensuremath{\,\mu\mathrm{m}}}
\newcommand{\sbunit}{\ensuremath{\,\mathrm{mag~arcsec^{-2}}}}

\newcommand{\ld}{\mbox{$\lambda$}}
\newcommand{\ldld}{\mbox{$\lambda\lambda$}}

\newcommand{\scinot}[2]{${#1}\times 10^{#2}$}

\newcommand{\ddlr}{\ensuremath{\mbox{$d_{\mathrm{DLR}}$}}}

\newcommand{\bose}[2][]{#2}   %

\begin{document}

\title{The Type Ia Supernova 2021hem: A 2003fg-like Event in an Apparently Hostless Environment}
\titlerunning{The apparently hostless \sn}

\authorrunning{Bose, Stritzinger et al.}
\author{
Subhash Bose\inst{\ref{aarhus}}\orcid{0000-0003-3529-3854}
\and 
M. D. Stritzinger\inst{\ref{aarhus}}\orcid{0000-0002-5571-1833}
\and 
A. Malmgaard\inst{\ref{aarhus}}
\and 
C. J. Miller\inst{\ref{michigan}}\orcid{0000-0002-1365-0841}
\and 
N. Elias-Rosa\inst{\ref{padua}}\orcid{0000-0002-1381-9125}
\and 
J. P. U. Fynbo\inst{\ref{copenhagen}}\orcid{0000-0002-8149-8298}
\and 
C.~Ashall\inst{\ref{uni:hawaii}}\orcid{0000-0002-5221-7557}
\and 
C. R. Burns\inst{\ref{uni:OCIW}}\orcid{0000-0002-5221-7557}
\and 
J.~M.~DerKacy\inst{\ref{baltimore}}\orcid{0000-0002-7566-6080}
\and 
L. Galbany\inst{\ref{barcelona1},\ref{barcelona2}}\orcid{0000-0002-1296-6887}
\and C. P. Guti\'errez\inst{\ref{barcelona2},\ref{barcelona1}}\orcid{0000-0003-2375-2064}
\and 
W. B. Hoogendam\inst{\ref{uni:hawaii}}\orcid{0000-0003-3953-9532}
\and
E. Y. Hsiao\inst{\ref{uni:fsu}}\orcid{0000-0003-3953-9532}
\and
E.~A.~M.~Jensen\inst{\ref{uni:Chalmers}}\orcid{0000-0003-3197-3430}
\and 
K.~Medler\inst{\ref{uni:hawaii}}\orcid{0000-0001-7186-105X}
\and 
Alaa Alburai\inst{\ref{barcelona1},\ref{barcelona2}}\orcid{0009-0007-2731-5562}
\and 
J. Anderson\inst{\ref{uni:ESO}}\orcid{0000-0003-0227-3451}
\and 
E. Baron\inst{\ref{uni:baron1}}\orcid{0000-0001-5393-1608}
\and 
J.~Duarte\inst{\ref{Lisboa}}\orcid{0000-0003-3953-9532}
\and 
M.~Gromadzki\inst{\ref{uni:warshaw}}\orcid{0000-0002-1650-1518}
\and 
C. Inserra\inst{\ref{cardiff}}\orcid{0000-0002-3968-4409}
\and
P. A. Mazzali\inst{\ref{uni:LJM}}\orcid{0000-0001-6876-8284}
\and
T.~E.~Müller-Bravo\inst{\ref{dublin},\ref{ICEN}}\orcid{0000-0003-3939-7167}
\and 
P.~Lundqvist\inst{\ref{uni:stockholm}\orcid{https://orcid.org/}}
\and 
A. Reguitti\inst{\ref{padua}}\orcid{0000-0003-4254-2724}
\and 
I. Salmaso\inst{\ref{inaf-napoli},\ref{padua}}\orcid{0000-0003-1450-0869}
\and 
D. J. Sand\inst{\ref{tucson}}\orcid{0000-0003-4102-380X}
\and 
G. Valerin\inst{\ref{padua}}\orcid{0000-0002-3334-4585}
}

\institute{
	Department of Physics and Astronomy, Aarhus University, Ny Munkegade 120, DK-8000 Aarhus C, Denmark\label{aarhus} (\email{email@subhashbose.com, max@phys.au.dk})
	\and 
	Department of Astronomy, University of Michigan, Ann Arbor, MI 48109, USA\label{michigan}
	\and 
    INAF – Osservatorio Astronomico di Padova, Vicolo dell’Osservatorio 5, I-35122 Padova, Italy\label{padua}
    \and 
    Cosmic DAWN Center, Niels Bohr Institute, University of Copenhagen, Lyngbyvej 2, 2100 Copenhagen \O, Denmark\label{copenhagen}
    \and 
    Institute for Astronomy, University of Hawai'i, 2680 Woodlawn Drive, Honolulu HI 96822, USA\label{uni:hawaii}
    \and 
    Observatories of the Carnegie Institution for Science, 813 Santa Barbara St, Pasadena, CA, 91101, USA\label{uni:OCIW}
    \and 
    Space Telescope Science Institute, 3700 San Martin Drive, Baltimore, MD 21218-2410, USA\label{baltimore}
    \and 
	Institute of Space Sciences (ICE, CSIC), Campus UAB, Carrer de Can Magrans, s/n, E-08193 Barcelona, Spain\label{barcelona1}
	\and
	Institut d’Estudis Espacials de Catalunya (IEEC), E-08034 Barcelona, Spain\label{barcelona2} 
    \and Department of Physics, Florida State University, 77 Chieftan Way, Tallahassee, FL 32306, USA\label{uni:fsu}
    \and 
    Department of Physics, Chalmers University of Technology, SE-412 96, Göteborg, Sweden\label{uni:Chalmers}
    \and 
    European Southern Observatory, Alonso de C\'ordova 3107, Casilla 19, Santiago, Chile\label{uni:ESO}
    \and 
    Planetary Science Institute, 1700 East Fort Lowell Road, Suite 106, Tucson, AZ 85719-2395, USA\label{uni:baron1}
    \and 
    CENTRA, Instituto Superior T\'ecnico, Universidade de Lisboa, Av. Rovisco Pais 1, 1049-001 Lisboa, Portugal\label{Lisboa}
    \and 
     Astronomical Observatory, University of Warsaw, Al. Ujazdowskie 4, 00-478 Warszawa, Poland\label{uni:warshaw}
    \and 
     Cardiff Hub for Astrophysics Research and Technology, School of Physics \& Astronomy, Cardiff University, Queens Buildings, The Parade, Cardiff, CF24 3AA, UK\label{cardiff}
     \and 
     Astrophysics Research Institute, Liverpool John Moores University, IC2, Liverpool Science Park, 146 Brownlow Hill, Liverpool L3 5RF, UK\label{uni:LJM}
     \and 
     School of Physics, Trinity College Dublin, The University of Dublin, Dublin 2, Ireland\label{dublin}
	\and 
    Instituto de Ciencias Exactas y Naturales (ICEN), Universidad Arturo Prat, Chile\label{ICEN}
	\and 
    The Oskar Klein Centre, Department of Physics, Stockholm University, AlbaNova, 10691 Stockholm, Sweden\label{uni:stockholm}
    \and 
    INAF - Osservatorio Astronomico di Capodimonte, Salita Moiariello 16, 80131 Napoli, Italy\label{inaf-napoli}
    \and 
	Steward Observatory, University of Arizona, 933 North Cherry Avenue, Tucson, AZ 85721-0065, USA\label{tucson}
}

\date{Received 10 November 2025 / Accepted 24 November 2025.}

\abstract{
We report observations of a Type Ia supernova 2021hem, discovered within 48 hours of last non-detection, and located in an apparently hostless environment.
With a peak absolute $B$-band magnitude of $M_{B,\max} = -19.96 \pm 0.29$\,mag, \sn\ lies at the luminous end of the SN~Ia distribution. Its near-infrared and $i$-band light curves lack the secondary maximum, which is otherwise ubiquitous to normal and 1991T-like SNe~Ia. Instead, these properties make \sn\ closely resemble 2003fg-like events. The slowly evolving light curves (characterized by $\Delta m_{15}(B) = 1.02 \pm 0.02$ mag; $s_{BV} = 0.94 \pm 0.05$), and the earliest spectrum showing \ion{C}{ii}~\ld6580 and \ld7235 absorption lines, further support this classification. Other spectroscopic features, including \Siii\ line diagnostics, resemble normal SNe~Ia.
A fit of a ``fireball'' model to the early-time light curves yields a time of first light of $t_{\rm first} = -16.43^{+0.45}_{-0.38}$\,days relative to $B$-band maximum. The first photometric detection occurs $1.51^{+0.45}_{-0.38}$\,days before the onset of fireball-like flux rise. This early emission, together with the intrinsic $(g - r)_0$ color, is inconsistent with circumstellar or companion interaction. Instead, shallow \nickel\ mixing or an asymmetric \nickel\ distribution offers a plausible explanation for the delayed onset of the fireball flux rise, while a double-detonation scenario with a thin helium shell remains a less likely alternative.
Notably, \sn\ represents the fifth known 2003fg-like SN that has early-time activity or excess flux emission. 
The estimated mass of radioactive \nickel\ synthesized in \sn\ is $1.00 \pm 0.09$\,\msun. 
Deep GTC imaging obtained 2.5\,years after the explosion, with an estimated limiting magnitude of $m_{lim,r} = 24.4$\,mag and a surface-brightness limit of $\mu_{lim,r} = 26.3$\,\sbunit, reveals no coincident host, thereby ruling out most faint dwarf and ultra-diffuse galaxies (UDGs). 
Alternatively, if the nearest plausible AGN host galaxy located at a projected distance of 104\,\kpc\ is assumed, the progenitor would need to be a hyper-velocity star ejected at $\sim2200$\,\kms\ from the host \bose{by AGN interaction}.  A faint diffuse feature $\approx6$\kpc\ from the SN site has also been detected in the GTC image, with its surface brightness within the limits of UDGs. However, it is unclear whether it is a galaxy and is associated with \sn. Considering its large normalized directional light distance (\ddlr\ $\sim3-4$) from the SN, and its unusual elongation, it is a candidate of low probability to be the host galaxy of \sn.
These results identify \sn\ as one of the strongest candidates for a hostless SN~Ia, underscoring the diversity of luminous, slowly evolving, 2003fg-like explosions and 
the wide range of environments in which they may occur.
}

\keywords{supernovae : general -- supernovae: individual: SN~2021hem}

\maketitle

\section{Introduction}

Type\,Ia supernovae (SNe\,Ia) are thermonuclear explosions of carbon–oxygen white dwarfs (WDs), generally thought to occur when the progenitor approaches the Chandrasekhar mass limit of $\approx~1.4$\,$M_{\odot}$ \citep{Nomoto1984,Hillebrandt2000}. Their relative uniformity in peak luminosity has made them indispensable as cosmological distance indicators \citep{Riess1998,Perlmutter1999}.
Beyond the population of normal SNe~Ia, several distinct subclasses have been identified, including the low-luminosity, red 1991bg-like events \citep[e.g.,][]{1992AJ....104.1543F,1993AJ....105..301L}, the overluminous, blue 1991T-like SNe \citep[e.g.,][]{1992ApJ...384L..15F,1992AJ....103.1632P,2024ApJS..273...16P}, and the exceptionally bright 2003fg-like explosions \citep{Howell2006}.
Additional peculiar subtypes have also been recognized, such as the 2002cx-like (commonly referred to as  SNe~Iax; \citealt{2003PASP..115..453L,2013ApJ...767...57F}) and the 2002es-like SNe \citep[e.g.,][]{2012ApJ...751..142G,Bose2025}.

Among the rarest subtypes of SNe~Ia are the 2003fg-like events \citep{Howell2006,2007ApJ...669L..17H}, historically referred to as ``super-Chandrasekhar'' SNe~Ia. These objects exhibit broad, slowly declining light curves and peak luminosities exceeding those of normal SNe~Ia with comparable decline rates. They also lack the secondary maximum in near-infrared (NIR) light curves that typically characterizes normal or luminous 1991T-like SNe, while often displaying several distinctive features. These include: low expansion velocities, a prominent \ion{C}{ii}~$\lambda$6580 absorption feature near maximum light, and early light curves that show a short-lived flux excess, possibly arising from interaction with circumstellar material.  Furthermore, their relatively low expansion velocities, despite their high luminosities, are consistent with large ejecta masses \citep{Howell2006,2011MNRAS.412.2735T,Scalzo2010}. This  led to the interpretation that they originate from super-Chandrasekhar-mass WD progenitors. However, some 2003fg-like SNe have inferred ejecta masses within the Chandrasekhar-mass limit \citep[see][]{2014MNRAS.443.1663C,Lu2021}, suggesting a diversity in progenitor systems. 
The requirement for high ejecta masses in these events could also be satisfied by the merger of two near-Chandrasekhar-mass WDs \citep[e.g.,][]{2023MNRAS.521.1162D}.
Alternatively, CSM-interaction models have been invoked to explain the high luminosity in several 2003fg-like SNe \citep[e.g.,][]{2012MNRAS.427.2057H,2020ApJ...900..140H,2024A&A...687L..19N,2025MNRAS.542.2752B}.
Many 2003fg-like SNe also show a tendency to occur in low-mass, often metal-poor, dwarf host galaxies \citep{Childress2011,Khan2011}.  Although studies based on larger samples indicate that such events can arise in hosts spanning a wide range of masses \citep{2025A&A...694A..10D,2025MNRAS.542.2752B}.

 SNe~Ia occur in  a wide range of galactic environments, extending from star-forming spirals to quiescent ellipticals, and even within dense galaxy clusters, where intracluster stars might serve as progenitors \citep{2006ApJ...648..868S,Lampeitl2010,Pan2014}. While the majority of SNe~Ia are clearly associated with galaxies, a small fraction appear to lack a  host-galaxy association.
Only a handful of confirmed SNe~Ia to date have been identified in the intracluster regions of galaxy clusters, located far from any cluster members and likely originating from the intracluster stellar population \citep{2003AJ....125.1087G,2011ApJ...729..142S,2015ApJ...807...83G}. In principle, ``hostless'' SNe could also arise from hyper-velocity stars that escaped their host galaxies prior to explosion. However, no confirmed examples of this scenario have been reported to date.
Many candidates labeled as hostless in imaging surveys may, in fact, be associated with low surface brightness dwarf galaxies that remain undetected at the survey’s imaging depth \citep[e.g.,][]{Strolger2002,2025ApJ...988..278S}. This underscores the need for deep imaging to robustly confirm truly hostless SNe.

The Type~Ia SN~2021hem (ZTF21aaqwjlz) provides an important case of this phenomenon.  The absence of a clear host galaxy or any nearby cluster in its field raises the question of whether the explosion was truly hostless or associated with a system below current detection limits.
This possibility connects naturally to the population of low-surface-brightness galaxies. Ultra-diffuse galaxies (UDGs), with effective radii of $r_{\rm eff} \gtrsim 1.5$\,kpc, having low stellar masses of $\sim10^8\,\msun$, and central surface brightnesses of $\sim24$--26.5 mag arcsec$^{-2}$, have been identified in clusters, groups, and lower-density environments \citep{vanDokkum2015,Koda2015,Munoz2015,Mihos2015,vanderBurg2016,Ramon2017,Janssens2017,2017A&A...607A..79V}. At even fainter extremes lie the ultra-faint dwarf galaxies (UFDGs), with luminosities as low as $M_V \sim -1$ to $-8$ and surface brightnesses $\mu_V \sim 27$--31 mag arcsec$^{-2}$, making them among the most diffuse stellar systems known \citep{Willman2005,Belokurov2007,McConnachie2012,DrlicaWagner2015,Torrealba2019,2019ARA&A..57..375S,Collins2022}. 
They are typically comparable in size to globular clusters, with radii ranging from 30 to 70 pc, although a few larger ones can extend to a few hundred parsecs. Unlike globular clusters, however, UFDs contain several orders of magnitude fewer stars and exhibit higher velocity dispersions of 2–10~\kms, making them among the most dark-matter dominated galactic systems \citep{2023A&A...679A...2R}.
Due to their low surface brightness and small sizes, UFDGs cannot be detected beyond $\sim$10\,Mpc, even with 10-m class telescopes under arcsecond seeing conditions. The most distant UFDG identified so far, found in HST imaging, lies at $\sim$19\,Mpc \citep{2017ApJ...835L..27L}. Nevertheless, owing to their extreme faintness, these galaxies would always remain plausible hosts for otherwise ``hostless'' supernovae.

In this paper, we report follow-up data on SN~2021hem, obtained from shortly after explosion through over a month beyond peak brightness, in an apparently hostless environment. This case illustrates the challenges of identifying the faintest hosts and probing the extremely low-luminosity galaxy population. Recent results from the ELEPHANT project  indicate that fewer than $\sim$2\% of extragalactic transients are potentially hostless \citep{Pessi2024}. This highlights the need for deeper surveys to identify such rare events with better certainty.

The paper is organized as follows. Sect.~\ref{sec:discovery} summarizes discovery and classification. Sect.~\ref{sec:observations} describes the photometric and spectroscopic data sets, reductions, deep late-time imaging, and reddening and redshift estimates. Sect.~\ref{sec:results} presents the light-curve analysis, spectroscopic diagnostics. Sect.~\ref{sec:hostgalaxy} details the search for an underlying host, deriving point-source and surface-brightness limits and noting a nearby diffuse feature. Finally, Sect.~\ref{sec:summary} concludes with a brief summary.

\section{Discovery and classification}
\label{sec:discovery}

\sn\ was identified by  ALeRCE (Automatic Learning for the Rapid Classification of Events) broker system with J2000 coordinates R.A. $= 16^{h}21^{m}16^{s}.01$ and Decl. $= +14^\circ33'09\farcs61$ \citep{Forster2021}.
This discovery was made possible with ALeRCE’s stamp classifier \citep{Carrasco-Davis2021}, utilizing data from the Zwicky Transient Facility (ZTF; \citealt{Bellm2019}) public data stream. According to the Transient Name Server (TNS) discovery certificate, the first detection of the object appeared in an $r_{ZTF}$-band image taken on 2021-03-24T10:05:20.00 UT and the last non-detection was obtained 45 hours earlier (i.e,  2021-03-22T11:33:32 UT). Making use of the ZTF forced photometry service \citep{Masci2023}, we recomputed photometry of the ZTF data stream which indicates  at the time of discovery of \sn\ had an $r_{ztf}$-band magnitude of 19.9, while the limiting magnitude of the non-detection image was 20.12 mag. 

An optical spectrum obtained  with the 6.5-m MMT Observatory equipped with Binospec on 2021-04-01T09:33:20 allowed \citet{Terreran2021} to classify SN~2021hem as a SN~Ia. They also reported a redshift of $z = 0.035$,  presumably inferred from spectral template comparisons.

\begin{figure}[!t]
    \centering
	\includegraphics[width=1\linewidth]{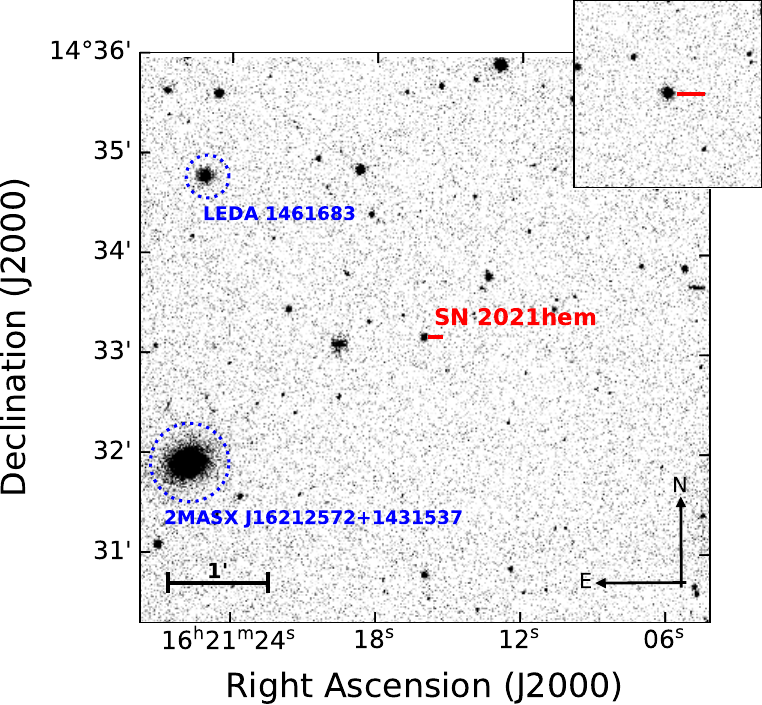}
	\caption{$r$-band image of \sn\ at peak brightness observed  with the Nordic Optical Telescope.  The location of \sn\ is positioned at the center in both panels. Two of the closest cataloged galaxies are marked in blue, both at the projected distance of roughly $\sim100\kpc$ (see Sect.~\ref{sec:arunawayburnoutnooneloved})}
	\label{fig:findingchart}
\end{figure}
Based on the initial  ZTF public stream data and demonstrated by the finding chart provided in Fig.~\ref{fig:findingchart}, \sn\ appears to lack an associated host galaxy. This motivated our team to conduct a followup campaign covering the early photospheric phase evolution.

\begin{figure}[!t]
\includegraphics[width=1.\linewidth]{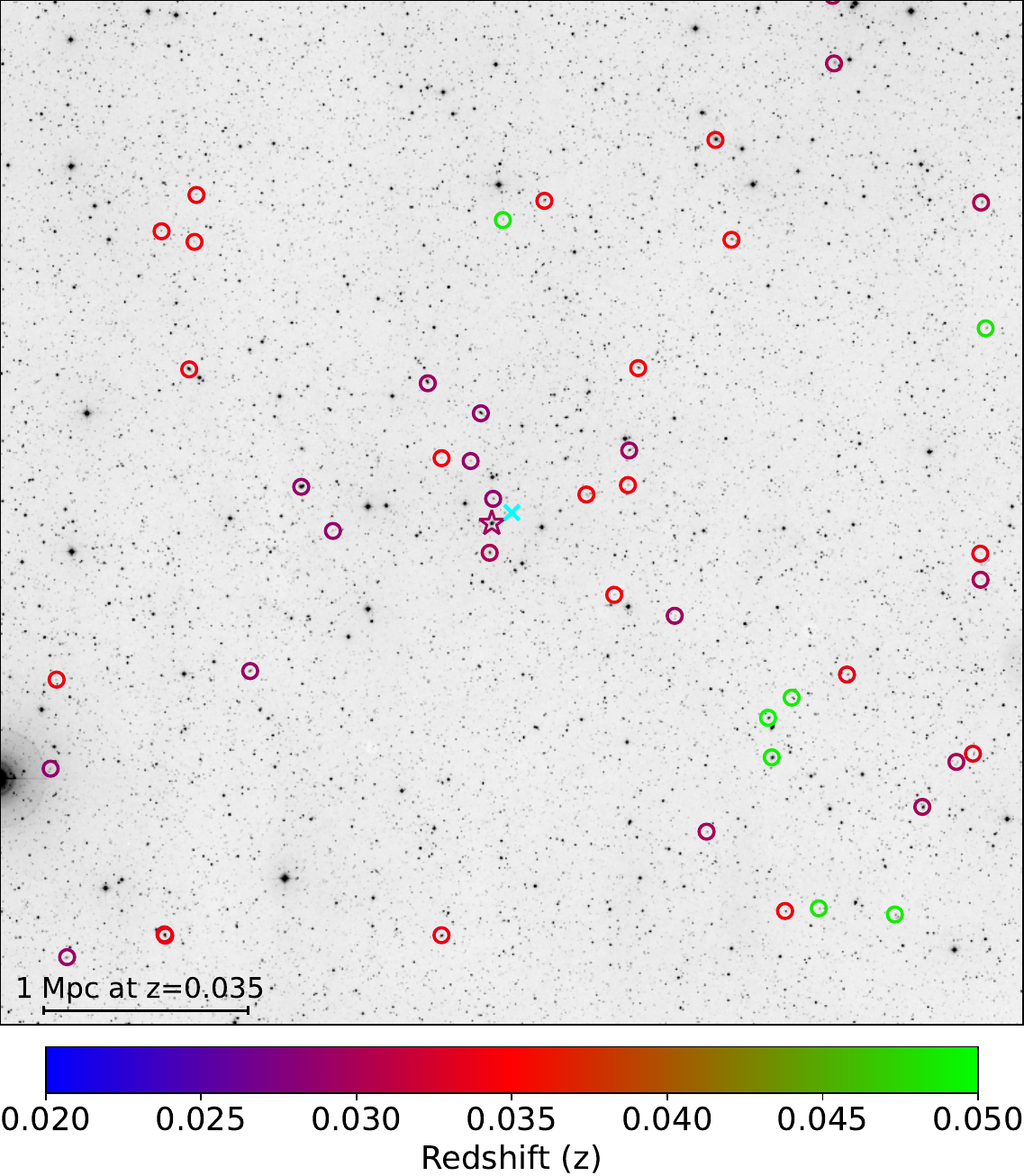}
\caption{Large-scale structure around the position of \sn\ (marked with a cyan cross) shown in a field of a size of 2$\times$2 square degrees DSS image. The colored circles indicate the galaxies with measured redshifts from SIMBAD catalog within redshift range of 0.02 to 0.05. The color-bar indicates the redshifts for each color. There appears to be an overdensity of galaxies with redshifts close to $z \sim 0.027$, $\sim0.035$ within a radius of 1\,Mpc from the position of the SN.
The open star symbol corresponds to the \bose{AGN galaxy 2MASX~J16212572{+}1431537,} which has the projected distance of 104\kpc\ from the SN.}
\label{fig:environment}
\end{figure}

\section{Observations}
\label{sec:observations}

\subsection{SN data}

Our characterization of  the photometric properties of \sn\ are based on a collection of photospheric phase optical $BgVri$- and near-IR $JH$-band photometry.  
During the course of the followup campaign, 22 epochs of optical imaging were obtained with the Las Cumbres Observatory's Global Network of 1-m robotic telescopes and a single epoch with the Nordic Optical Telescope (NOT) equipped with the Alhambra Faint Object Spectrograph and Camera (ALFOSC). Two epochs of the  $JH$-band imaging were also obtained with the NOT, equipped with the NOT near-infrared Camera and spectrograph (+ NOTcam), while an additional epoch of $JH$-band imaging was obtained with the New Technology Telescope (NTT) equipped with  the Son OF ISAAC (Sofi).

The optical and NIR imaging data sets were reduced following standard procedures.   PSF photometry was computed for the SN using the FLOWS\footnote{https://flows.phys.au.dk}  photometry pipeline, with nightly zeropoints computed relative to stars in the field with ATLAS Refcat2  catalog magnitudes \citep{Tonry2018b}.

 Our Las Cumbres photometry is complemented by the ZTF public data stream, and Pan-STARRS photometry through Young Supernova Experiment \citep[YSE;][]{2021ApJ...908..143J}. The ZTF forced photometry service was used to compute up to date photometry and we ensured the implementation of the baseline correction \citep[see][for details]{Masci2023}. This data provides high cadence photometry beginning from around the time of first light (hereafter $t_{first}$; see Sect.~\ref{sec:tfirst}) and extending though the primary maximum. Finally, for completeness,  we also ingested $o$- and $c$-band forced photometry from The Atlas Project \citep{Tonry2018a,Shingles2021}. The photometry measurements are listed in Tables~\ref{tab:photsn}-\ref{tab:photsnNIR}.

\begin{table*}
\centering
\caption{Journal of Spectroscopic Observations.\label{tab:specjor}}
\begin{tabular}{llrllrlr}
\hline
\hline 
Date & Julian Date & Phase\tablefootmark{a} & Telescope & Instrument  & Integration (s) & $z$ & rlap\tablefootmark{b}\\
\hline
\multicolumn{8}{c}{\bf Optical}\\
2021 April 01\tablefootmark{c}  & 2459305.90  & $-$10.19  & MMT  & BINOSPEC & 300  & $0.030\pm0.006$ & 9.6 \\ 
2021 April 10  & 2459314.57 & $-$1.82    & NOT  & ALFOSC   & 900  & $0.037\pm0.005$ & 12.1\\
2021 April 13  & 2459317.90 & $+$1.39    & APO 3.5-m & DIS   & 1800  & $0.036\pm0.006$ & 14.5 \\
2021 April 16  & 2459320.56 & $+$3.96    & NOT  & ALFOSC   & 900  & $0.035\pm0.003$ & 18.1\\
2021 April 20  & 2459324.62 & $+$7.87    & NOT  & ALFOSC   & 900  & $0.044\pm0.004$ & 16.0\\
2021 May 05    & 2459338.69 & $+$21.45   & NOT  & ALFOSC   & 900  & $0.035\pm0.004$ & 15.0\\
2021 May 16    & 2459351.50 & $+$33.81   & NOT  & ALFOSC   & 1800 & $0.035\pm0.003$ & 17.3\\
\hline 
\multicolumn{8}{c}{\bf NIR}\\
2021 April 11    & 2459315.91& $-0.53$   & IRTF  & SpeX   & 1079 & $\cdots$ & $\cdots$\\
\hline 
\end{tabular}
\tablefoot{\\
\tablefoottext{a}{Restframe days relative to the epoch of $B$-band maximum, i.e., \PeakEpoch.}
\tablefoottext{b}{Each redshift inferred by SNID carries a rlap parameter. This is a reliability metric that reflects both the strength of the spectral cross-correlation and the extent of wavelength overlap between the observed spectrum and the matched template. Higher rlap values correspond to better spectral matches, with values above $\sim 5$ generally considered significant for classification.}
\tablefoottext{c}{\citet{Terreran2021} reported a redshift of $z=0.035$ to the Transient Name Server along with their classification spectrum.}
}
\end{table*}

In addition to the  classification spectrum available on TNS (Transient Name Server), we obtained, in coordination with  NUTS2 (NOT Unbiased Transient Survey),  5 optical spectra with the NOT (+ ALFOSC), as well as a single spectrum with the APO 3.5-m (+ DIS, dual imaging spectrograph).
The NOT data were reduced following standard procedures using a software graphical user interface written in \texttt{pyraf}. Nightly sensitivity functions were computed based on the observations of at least one nightly spectrophometric flux standard. Similar reductions were performed using \texttt{IRAF} packages on the APO data obtained with both red and blue gratings. 
Complementing the optical spectroscopic time-series is a single NIR spectrum of SN~2021hem obtained with the NASA IRTF (+ SpecX) as part of The Hawaii Infrared Supernova Study \citep{2025arXiv250518507M} at around peak brightness.
A journal of spectroscopic  observations is provided in Table~\ref{tab:specjor}.

\subsection{Late-time deep imaging by GTC}

To characterize the location of \sn, we obtained a sequence of $r$-band images with 
the Gran Telescopio CANARIAS (GTC) equipped with OSIRIS (Optical System for Imaging and low-Intermediate-Resolution Integrated Spectroscopy), $\sim900$\,days after the explosion on 9 September 2023.  At this phase, the SN is expected to be around $ \sim27-28$\,mag assuming SN~2011fe like light curve, which is well below the detection limit of the image (see Sect.~\ref{sec:limmag} for the precise estimate of the image limiting magnitude).
In total, 7 individual images of 60 second integration time were obtained, along with  a single image with a 15 second integration time. 

The images were reduced using standard procedures with software based on the \texttt{photutils} package. Each image was bias-subtracted, flat-field corrected, and cleaned of cosmic rays. The images were then geometrically aligned and combined to produce a single science frame.
This image is used in Sect.~\ref{sec:limmag} to estimate the limiting magnitude of any potential underlying host galaxy.

\subsection{Reddening}
\label{sec:reddening}
The Milky Way  reddening  color excess  along the line-of-sight toward \sn\ is $E(B-V)_{MW} =0.0501 \pm  0.0034$\,mag  \citep{2011ApJ...737..103S}.
The post-peak $(B-V)_0$ color curve of \sn, after correcting for Milky way reddening is consistent with the intrinsic color-curve of the archetypal normal SN~2011fe as well as the subluminous SN~1991bg (see Fig.~\ref{fig:color_comp}), suggesting \sn\ suffered minimal host extinction. Further corroborating this, \sn\ spectra reveal no \Nai~D absorption at the redshift of the SN. The minimal host reddening is also consistent with the apparent hostless environment of the SN. 
Therefore, we adopt the previously stated Milky Way reddening value in our extinction correction for SN~2021hem. Assuming $R_V = 3.1$,  this corresponds to 
an $A^{tot}_V = 0.155\pm0.011$\,mag.

\subsection{Redshift of \sn}
\label{sec:redshift}
Using the Supernova Identification (\texttt{SNID}) code \citep{Tonry1979,Blondin2007}, we determined the redshift of \sn\ for each spectrum by comparing it to a library of supernova spectral templates. The individual redshift estimates, their associated uncertainties, and the corresponding rlap values -- which quantify the quality of the spectral matches -- are listed in Table~\ref{tab:specjor}. We computed a weighted mean redshift incorporating both the statistical uncertainties and the rlap values as quality weights, yielding a redshift of $z=0.0363\pm0.0044$.
To account for potential systematic uncertainties in the \texttt{SNID}-based redshift estimates, arising from template mismatch, phase coverage limitations, and spectral quality, a systematic error floor of $\Delta z_{sys} = 0.002$ is adopted \citep{Blondin2007,Foley2009}, which was added in quadrature to the statistical uncertainty. This yields a redshift estimate of $z = 0.0363 \pm 0.0049$. This value is adopted in the following to apply redshift corrections and to compute the corresponding luminosity distance of $D_L = 165\pm22$~Mpc assuming a standard Planck cosmology \citep{2016A&A...594A..13P}, and the equivalent distance modulus of $36.09\pm0.29$\,mag.
We note the potential bias in SNID template matching for a low velocity SN, which we discuss in detail in Sect.~\ref{sec:redshiftbias}, and conclude that \sn\ is a normal velocity SN~Ia and unlikely there is any significant bias in the determined redshift.

As a consistency check, we examine the redshift distribution of galaxies within the field of view of \sn, as shown in Fig.~\ref{fig:environment}. The distribution reveals over-densities at redshifts of $z \sim 0.027$, $\sim0.035$, as well as  a weaker concentration at $\sim0.050$, suggesting the presence of three major galaxy groups along the line-of-sight to \sn. Given the redshift determined from SNID template matching, $z=0.0363$, \sn\ is plausibly associated with the galaxy group at $z\sim0.035$.

\begin{figure*}
    \centering
	\includegraphics[height=12cm]{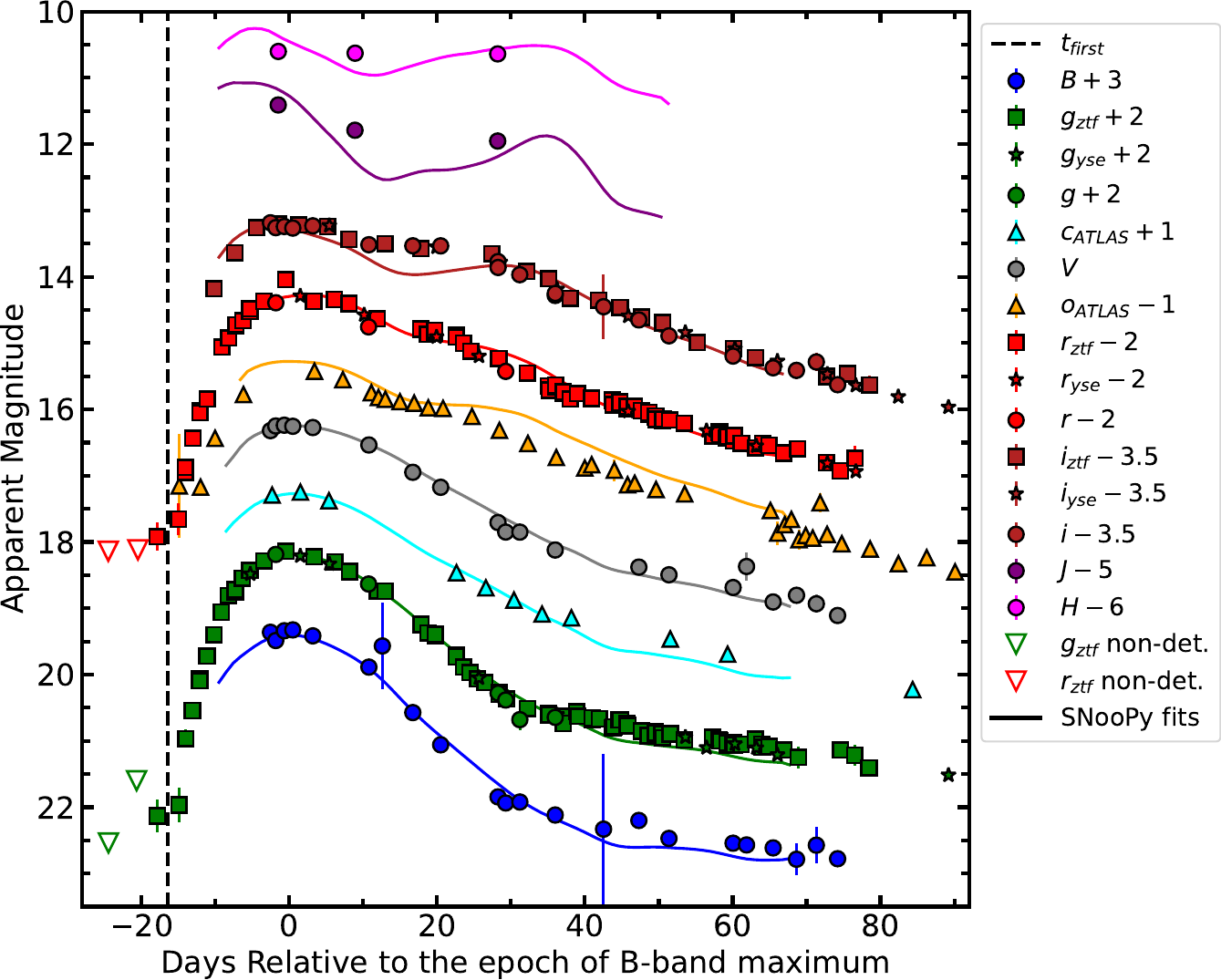}
	\caption{Optical and NIR light curves of \sn, shown alongside the best-fit \texttt{SNooPy} template light curves using the `\texttt{max\_model}'. 
    Our own observations are shown with filled circles, while photometry from ZTF, YSE, and ATLAS surveys are shown in different symbols. 
    The \texttt{SNooPy} template fits are overplotted as solid lines; those in the $gri$~bands are derived specifically from fits to the ZTF photometry. The indicated offsets have been added to the light curves for clarity. The vertical dashed line corresponds to the estimated ``time of first light'' by fitting a ``fireball'' model (see Sect.~\ref{sec:tfirst}). 
    }
	\label{fig:photometry}
\end{figure*}

\section{Results}
\label{sec:results}
\subsection{Photometry}

\subsubsection{Light-curve properties and template fitting}
\label{sec:lcfits}

The optical and NIR light curves of SN~2021hem, plotted relative to the epoch of $B$-band maximum, are shown in Fig.~\ref{fig:photometry}. The $B$-band maximum is estimated to have occurred on  $\rm\PeakEpoch \pm 0.31$, based on a direct Gaussian Process spline fit to the $B$-band light curve. This has been used as the reference epoch throughout the paper.

Using spline fits, the estimated peak apparent magnitude is $m_{B,max}=16.34\pm0.01$\,mag, and the post-peak light curve decline is estimated to be $\dmb=1.02\pm0.02$\,mag. After correcting for Milky Way reddening, computed K-correction of $-0.005$\,mag, and luminosity distance for the adopted redshift of  $z=0.0363\pm0.0049$, the peak absolute magnitude of \sn\ is $M_{B,max}=-19.96\pm0.29$.
On constructing $(B-V)_0$ intrinsic color-curve with interpolated spline fits after correcting for reddening, the color-curve peaks at $28.3\pm1.1$\,day, which corresponds to color-stretch parameter \citep{2014ApJ...789...32B} $\sbv=0.94\pm0.05$.

In Fig.~\ref{fig:photometry}, the observed light curve data points are overlaid with template fits generated using the \texttt{SNooPy} package \citep{Burns2011}, adopting the `\texttt{max\_model}' formalism.
The `\texttt{max\_model}' in \texttt{SNooPy} fits the peak magnitude in each band directly by treating the maximum light in each filter as free parameters, while using the same template-driven time evolution (based on $\Delta m_{15}$) and applying $K$-corrections and Milky Way extinction consistent with the `\texttt{EBV\_model}' framework.
Unlike models that impose an explicit reddening law and intrinsic colors to simultaneously fit the multi-band light curves, the max\_model approach avoids specifying host-galaxy extinction and instead focuses on extracting peak brightnesses across all observed filters \citep[see also][their Sect.~3 for details]{Stritzinger2010}.

The `max\_model'  provides fit parameters of the time of maximum as ${\rm JD}~2459316.74\pm0.36$ and a color-stretch parameter $s_{BV} = 1.19\pm0.04$.
Table~\ref{tab:LCparameters} provides results from direct GP splines to the optical light curves, and the estimated peak apparent magnitudes of the optical/NIR light curves estimated from the \texttt{SNooPy} `\texttt{max\_model}' fits, which include corrections for Milky Way reddening and computed K-correction.
Adopting a redshift of 
$z=0.0363\pm0.0049$ for SN~2021hem (see Sect.~\ref{sec:redshift}), the peak apparent magnitudes computed by the `\texttt{max\_model}' indicate that it reached an absolute  $B$-band peak magnitude of
$M_B = -19.90\pm0.29$ mag, roughly half a magnitude brighter than typical SNe~Ia of similar color-stretch values. %

\begin{table}
\tiny
\caption{Light curve fits.\label{tab:LCparameters}}
\begin{tabular}{llll}
\hline
\hline 
Filter & JD peak & Peak app. Mag. & $\Delta m_{15}$\,(mag)\\
\hline
\multicolumn{4}{c}{\bf Gaussian Spline fits}\\
\hline 
$B$ & $2459316.46\pm0.31$ & $16.338\pm0.009$ & $1.036 \pm0.051$  \\
$g_{ztf}$ & $2459316.60\pm0.26$ & $16.159\pm0.014$ & $0.823\pm0.032$\\
$V$ & $2459317.39\pm0.37$ & $16.250\pm0.010$ & $0.607\pm0.034$ \\
$r_{ztf}$ & $2459316.07\pm0.31$ & $16.251\pm0.016$ & $0.469\pm0.024$ \\
$i_{ztf}$ & $2459315.69\pm0.32$ & $16.698\pm0.016$ & $0.379\pm0.028$ \\
\hline
\hline 
\multicolumn{4}{c}{\bf SNooPy `max\_model fit'\tablefootmark{a}}\\
\hline 
$B$       & $\cdots$  & $16.184\pm0.011\pm0.012$  & $\cdots$ \\
$g_{ztf}$ & $\cdots$  & $16.087 \pm 0.013 \pm 0.014$           & $\cdots$\\
$V$       & $\cdots$  & $16.115\pm0.015\pm0.019$  & $\cdots$ \\
$r_{ztf}$ & $\cdots$  & $16.184\pm0.012\pm0.022$  & $\cdots$ \\
$i_{ztf}$ & $\cdots$  & $16.567\pm0.026\pm0.022$  & $\cdots$ \\
$J$       & $\cdots$  & $16.163\pm0.077\pm0.044$  &  $\cdots$ \\
$H$       & $\cdots$  & $16.339\pm0.062\pm0.055$ & $\cdots$\\
\hline 
\end{tabular}
\tablefoottext{a}{Peak magnitudes include K-correction and extinction correction. Uncertainties correspond to fit and systematic errors.}
\end{table}

Inspection of the light curves also reveals several notable features suggesting SN~2021hem is a peculiar object. 
First, the flux evolution from the discovery epoch to the next observation shows a short plateau in the early light-curve and deviates markedly from a power-law rise, which is analyzed in detail in Sect.~\ref{sec:tfirst}. 
Second, the $i$-, $J$-, and $H$-band light curves lack secondary maxima and appear to be unusually flat compared to what expected from a typical SNe~Ia or luminous 1991T-like SNe~Ia. This is clearly reflected in the significant deviations between \sn\ and the \texttt{SNooPy} template light curve fits in these bands, which are drawn from normal SNe~Ia based on \dmb. Additionally, the NIR bands are nearly as bright as the optical, in contrast to normal SNe~Ia, where NIR absolute magnitudes are typically about one magnitude fainter. Taken together, these photometric peculiarities of \sn, along with high peak luminosity, makes it resemble other 2003fg-like SNe~Ia.

\begin{figure}[!t]
	\centering
        \includegraphics[width=1.\linewidth]{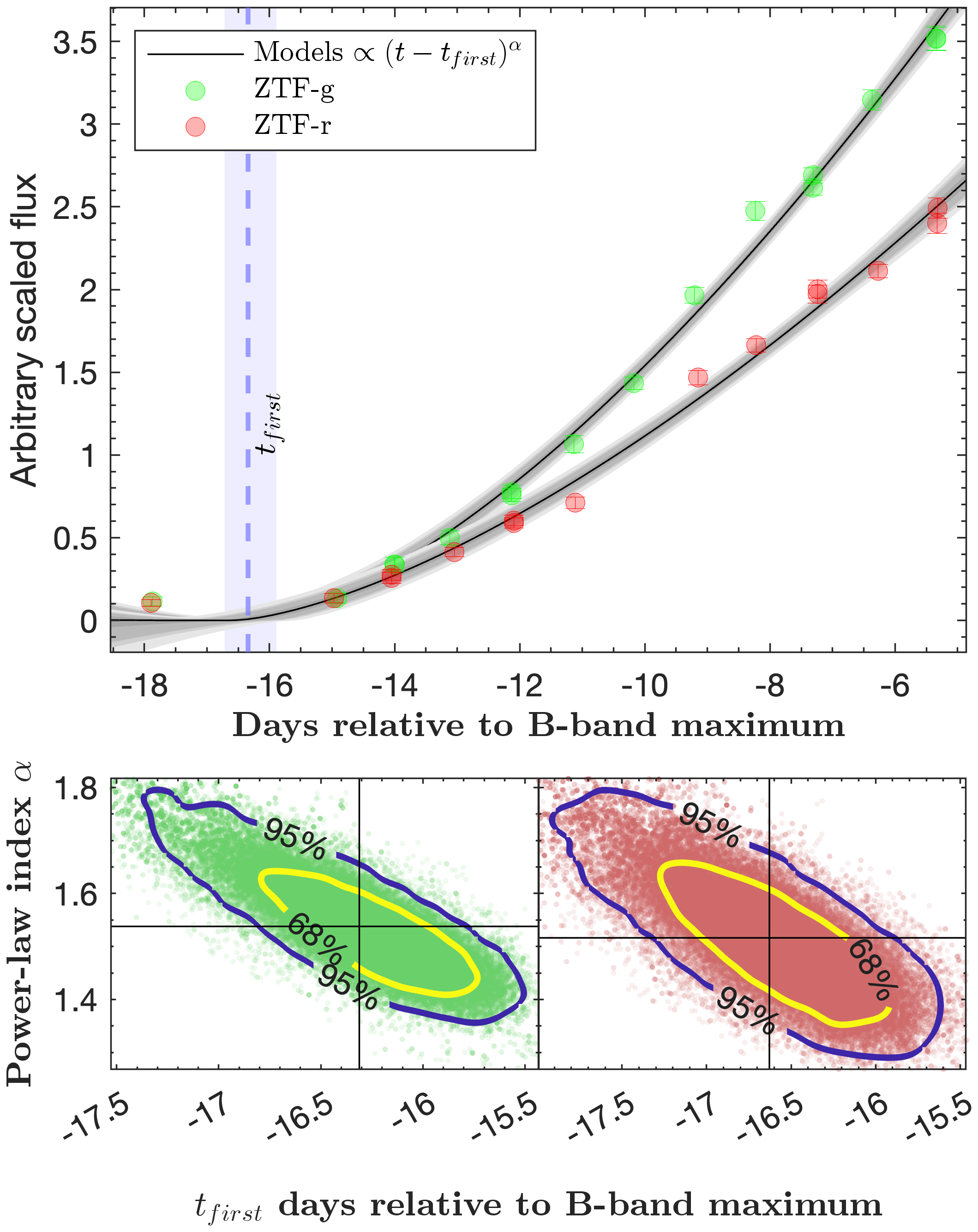}
	\caption{MCMC power-law model fit to ZTF $g$ and $r$-band fluxes to estimate the time of first-light $t_{first}$. The fitting is limited to the first 12 days after $t_{first}$, during which the light curve clearly follows a power-law rise. Power-law model fits are performed separately on the ZTF $g$- and $r$-band light curves to determine $t_{first}$ and $\alpha$ independently for each band. In the top panel, the $g$- and $r$-band fluxes are shown with arbitrary and distinct scales for visual clarity. The inferred (weighted mean) $t_{first}$ is shown by a vertical dashed line, with the shaded region indicating the uncertainty limits. In the bottom panels, the vertical and horizontal lines indicate the $t_{first}$ and $\alpha$ values, as inferred from each MCMC fit.}
\label{fig:tfirst}
\end{figure}

\citet{Ashall2020} proposed a photometric classification scheme for SNe~Ia based on the time difference between the $i$- and $B$-band maxima ($t_{\max}^{i-B}$) and the color-stretch parameter $s_{BV}$. Using the $B$- and $i$-band peak dates listed in Table~\ref{tab:LCparameters}, we derive a rest-frame corrected value of $t_{\max}^{i-B} = -0.77 \pm 0.45$~days for \sn. Combined with the previously estimated $s_{BV}$ range of 0.94–1.19, obtained from both direct $(B-V)_0$ color measurements and \texttt{SNooPy} fits, this places \sn\ among normal and luminous 1991T-like SNe~Ia in the \citet{Ashall2020} classification diagram. In contrast, 2003fg-like subclasses occupy the region with $t_{\max}^{i-B} > 0$ in the same diagram \citep{Ashall2021}. Therefore, \sn\ lies marginally outside the parameter space of 2003fg-like SNe sample that populate the \citet{Ashall2020,Ashall2021} classification diagram.

\subsubsection{Estimating the time of first light, $t_{first}$}
\label{sec:tfirst}

We fit the rising portions of the ZTF $g$- and $r$-band light curves individually using an MCMC power-law model (see Fig.~\ref{fig:tfirst}), following the methodology of \citet{Bose2025}.
The time of first light, $t_{first}$, defined as the point when the SN luminosity begins a power-law rise, is estimated to be $-16.63^{+0.71}_{-0.64}$~days relative to the epoch of $B$-band maximum, with a corresponding power-law index  $\alpha$ of $1.52^{+0.14}_{-0.17}$ form the ZTF $r$-band light curve. Similarly, using the ZTF $g$-band light curve, the estimated parameters are  $t_{first}=-16.31^{+0.58}_{-0.48}$ and $\alpha=1.54^{+0.10}_{-0.13}$. 
The parameter estimates from the individual light curves are highly consistent, and we adopt their weighted mean as $t_{first}=-16.43^{+0.45}_{-0.38}$ and $\alpha=1.53^{+0.08}_{-0.10}$.
This time of first light corresponds to $\rm JD~2459299.43^{+0.45}_{-0.38}$.

Interestingly, this epoch of $t_{first}$ occurs $1.51^{+0.45}_{-0.38}$ days after the initial detection (JD $\approx 2459297.92$) in the ZTF images. 
In fact, we re-ran the MCMC fit excluding the detection epoch, and found the epoch of first detection is outside 99.9\% confidence level of $t_{first}$ estimation.   Visual inspection of both the $r$- and $g$-band ZTF images confirms the presence of a source at this time, as shown in Fig.~\ref{fig:photometry}, marking the first detection in both the ZTF $g$ and $r$ bands at $\sim20$\,mags.
This suggests that SN~2021hem emitted detectable radiation several days before its light curve began to follow the expected fireball expansion model. 
 
The early SN atmosphere ($\lesssim7$\,day) can be well approximated as an optically thick ``fireball" expanding homologously, and would result in a power-law flux rise ($\propto t^\alpha$). Assuming a constant temperature would imply a power-law index of 2. Any variation in the power-law index would imply a deviation of energy transport within the ejecta, primarily due to non-uniformity of \nickel\ mixing. Studies implementing the fireball model with early SNe~Ia light curves have shown significant deviations from the classical fireball model power-law index of 2. A study on a sample of early TESS SNe~Ia light curves \citep{2023ApJ...956..108F} found a mean power-law index of $1.93 \pm 0.57$ with rise-time of $15.7 \pm 3.5$\,days, whereas another study using SNe~Ia in PTF and LSQ survey data \citep{2015MNRAS.446.3895F} found a mean power-law index of $2.44\pm0.13$. In the case of SN~2020qxp/ASASSN-20jq, \citet{Bose2025} found a power-law index, which is almost exactly the same as the classical fireball model's value of 2, indicating well-mixed \nickel\ in the ejecta, which is also consistent with their non-LTE nebular-phase spectroscopic model  that required macroscopic mixing to match observations. %

Although the power-law index of $~1.53$ for \sn\ lies within the range found in previous studies, it is on the lower side of the distribution and is significantly below the classical value of 2. The low power-law index of \sn\ possibly implies the SN had a shallow \nickel\ mixing, where the \nickel\ is confined deeper in the ejecta, thus the early light-curve rises more slowly because photons take longer to diffuse to the photosphere. 

As discussed above, the first photometric detections of \sn\ were made several days prior to the onset of the light curve’s power-law rise. One explanation for this pre-fireball emission is shallow \nickel\ mixing, where radioactive heating occurs deeper in the ejecta. This results in a ‘dark phase,’ during which $t_{first}$ is delayed by a few days after the explosion \citep[see, e.g.,][]{2013ApJ...769...67P, 2014ApJ...784...85P}. This interpretation of the dark phase is consistent with the evidence for shallow \nickel\ mixing discussed earlier in the context of the low power-law index. Another explanation for the delayed flux rise could be asymmetric distribution of \nickel\, where the majority of \nickel\ is on the far side of the explosion. This would appear to have the same effect as deeply embedded \nickel\ leading to a dark phase. 
An alternative interpretation, proposed for MUSSES1604D \citep{2017Natur.550...80J}, which displayed an early $\approx1$ day plateau in its $g$-band light curve, involves a white dwarf undergoing a surface helium detonation, also referred to as an edge-lit detonation \citep[e.g.,][]{2010A&A...514A..53F,2010ApJ...719.1067K}. This surface detonation subsequently triggers a secondary detonation in the white dwarf core with a few days of delay, which can naturally explain the delayed fireball flux rise.

This pre-fireball emission should not be confused with the early flux excess seen in many early SNe~Ia light curves, as the two phenomena are characteristically distinct. The early flux excess is typically attributed to the interaction of the SN ejecta with a non-degenerate companion star \citep{2010ApJ...708.1025K}, or with circumstellar material \citep[e.g.,][]{Piro2016,Maeda2023,2023MNRAS.522.6035M,2018ApJ...864L..35S}.
It is estimated that approximately $20-30\%$ of all SNe~Ia exhibit such early-excess features \citep{2022MNRAS.513.3035M, 2022MNRAS.512.1317D}. However, in certain subclasses of SNe~Ia -- such as the peculiar 2002es-like, 2003fg-like, and luminous 1991T-like events -- the occurrence rate of early-excess emission appears to be considerably higher \citep{2018ApJ...865..149J,2024ApJ...966..139H}.
Crucially, unlike these early-excess signatures, the pre-fireball emission observed in \sn\ likely arises from a distinct physical mechanism, as discussed earlier. An early flux excess due to interaction typically emerges within a few days of explosion and appears superimposed on the power-law rise of the early light curve. This is because the short-lived excess flux from the interaction and the underlying ``fireball’’ emission are both consequences of the expanding ejecta. In other words, the interaction can not occur without the ``fireball'' expansion. In contrast, the early emission seen in \sn\ occurs even before the onset of the fireball's power-law flux rise, clearly distinguishing it from the early excesses or bumps arising from interaction in other SNe~Ia.
Moreover, \cite{2017Natur.550...80J} argued, based on their models, that an early excess arising from interaction would inevitably show a blue color evolution, as was observed in the $(\bv)_0$ color of SN iPTF14atg. \bose[]{\cite{2020ApJ...902...48B} has also shown using ejecta-companion interaction models that the $(g-r)_0$ colors at the time of interaction is expected to be ${\sim}-0.5$\,mag.} In the case of \sn\ the $(g-r)_0$ color of  $0.16\pm0.33$\,mag, computed from first epoch of $g$ and $r$-band light curves, does not show any blue excess. This color is broadly consistent with, or perhaps marginally redder than, the early $(g-r)_0$ colors typically observed in normal SNe~Ia \citep{2020ApJ...902...48B}. Therefore, considering all these factors, the interaction scenario is disfavored for the pre-fireball emission seen in \sn.  

Regardless of the origin, this early activity, which can be described as early flux excess or pre-fireball flux emission as in \sn, appears to be a ubiquitous feature among 2003fg-like SNe. In addition to \sn, other 2003fg-like supernovae with early-time observations and adequate photometric cadence -- SNe 2020hvf, 2021zny, 2022ilv, and 2021qvo \citep{2021ApJ...923L...8J,2024ApJ...966..139H, 2025arXiv250813263A} -- have all exhibited an early bump in their light curves. %

\begin{figure}
\includegraphics[width=1\linewidth]{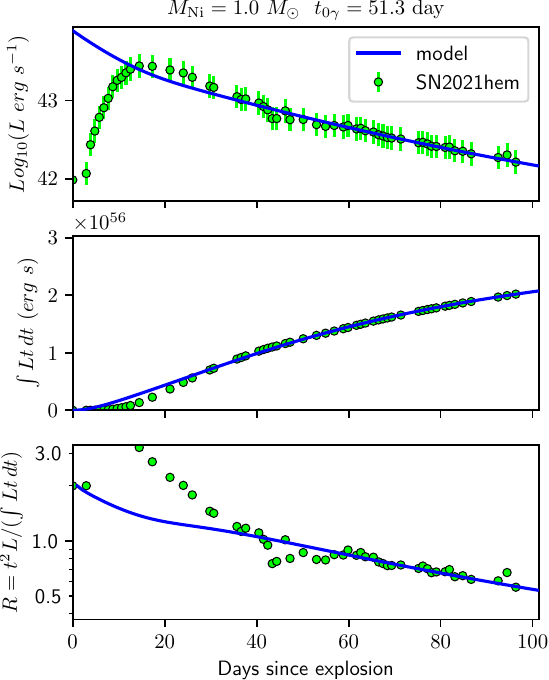}
\caption{UVOIR bolometric light curve of \sn\ fit with a radioactive \nickel\ decay model. The time of explosion is taken as the epoch of first detection, which is $\approx17.9$ days before the $B$-band maximum. Only data beyond 60 days are used for fitting the model. The UVOIR  light curve of \sn\ (top panel), its time-weighted integrated luminosity (middle panel), and the ratio $t^2L/\left(\int L t~dt\right)$ (bottom panel), which is defined independent of the \nickel\ mass are shown. 
}
\label{fig:nickelmodel}
\end{figure}

\subsubsection{Bolometric light curve and \nickel\ mass}

To estimate the quantity of \nickel\ synthesized during the disruption of \sn, we constructed and modeled a UVOIR (ultraviolet–optical–infrared) bolometric light curve.
Broad-band optical light curves were first interpolated using spline functions at coeval epochs. A total reddening correction of $E(B-V)_{\mathrm{tot}} = 0.05$~mag was applied (see Sect.~\ref{sec:reddening}). The bolometric flux was then estimated using the \texttt{SNooPy} package, with spectral templates from \citet{2007ApJ...663.1187H} color-matched to the reddening-corrected broad-band colors of \sn. To account for flux outside the spectral range of the templates, we (i) extended the spectral energy distribution (SED) bluewards of the atmospheric cutoff by linearly extrapolating to zero flux at 1000\,\AA, and (ii) extended the redwards end of the SED beyond the $i$-band using a Rayleigh–Jeans tail extending to 25,000\,\AA. The integrated flux was then converted to bolometric luminosity using the distance adopted in Sect.~\ref{sec:redshift}. The resulting UVOIR bolometric light curve is shown in the top panel of Fig.~\ref{fig:nickelmodel}.

To estimate the $^{56}$Ni mass, the UVOIR light curve of \sn\ was fit with an energy-deposition model corresponding to the radioactive decay chain $\nickel \rightarrow \cobalt \rightarrow \iron$. The best-fitting model is overplotted on the UVOIR light curve in the top panel of Fig.~\ref{fig:nickelmodel}. The model includes two free parameters: the synthesized \nickel\ mass $M_{\mathrm{Ni}}$ and the $\gamma$-ray trapping timescale $t_{0\gamma}$. We assumed complete deposition of the positron kinetic energy from \cobalt\ decay, while the fraction of trapped $\gamma$-ray energy was modeled as $\left[1 - \exp\left(-t_{0\gamma}^2 / t^2\right)\right]$.
At late epochs, when the contribution from positron energy becomes comparable to that of the deposited $\gamma$-ray energy, degeneracy arises between $M_{\mathrm{Ni}}$ and $t_{0\gamma}$. 
To address this point, we examined the time-weighted integral of the luminosity (middle panel of Fig.~\ref{fig:nickelmodel}). The result indicates that, at all times during the evolution of \sn, the energy budget can be fully explained by the \nickel\ decay chain \citep{Katz2013}, with no indication from the model fit of any additional sources of energy input.
 The bottom panel of Fig.~\ref{fig:nickelmodel} shows the ratio $t^2 L / \int L t\,dt$, which is defined to be independent of $M_{\mathrm{Ni}}$, and therefore allows for an independent determination of $t_{0\gamma}$ \citep{Katz2013}. The model was fit only using data obtained beyond 60~days after explosion, when the ejecta is expected to be nearly optically thin and the luminosity is almost solely powered by radioactive decay.
The best-fit parameters are $t_{0\gamma} = 51 \pm 3$~days and $M_{\mathrm{Ni}} = 1.0 \pm 0.1$~\msun. The inferred \nickel\ mass is significantly larger than those typically found in normal SNe~Ia \citep[e.g.,][]{2006A&A...450..241S, 2019MNRAS.483..628S,2020MNRAS.496.4517S}, but is consistent with larger values inferred for 2003fg-like SNe~Ia \citep[e.g.,][]{2006Natur.443..308H, 2007ApJ...669L..17H, 2012ApJ...756..191K,2019ApJ...880...35C}.

\begin{figure}
\includegraphics[width=\linewidth]{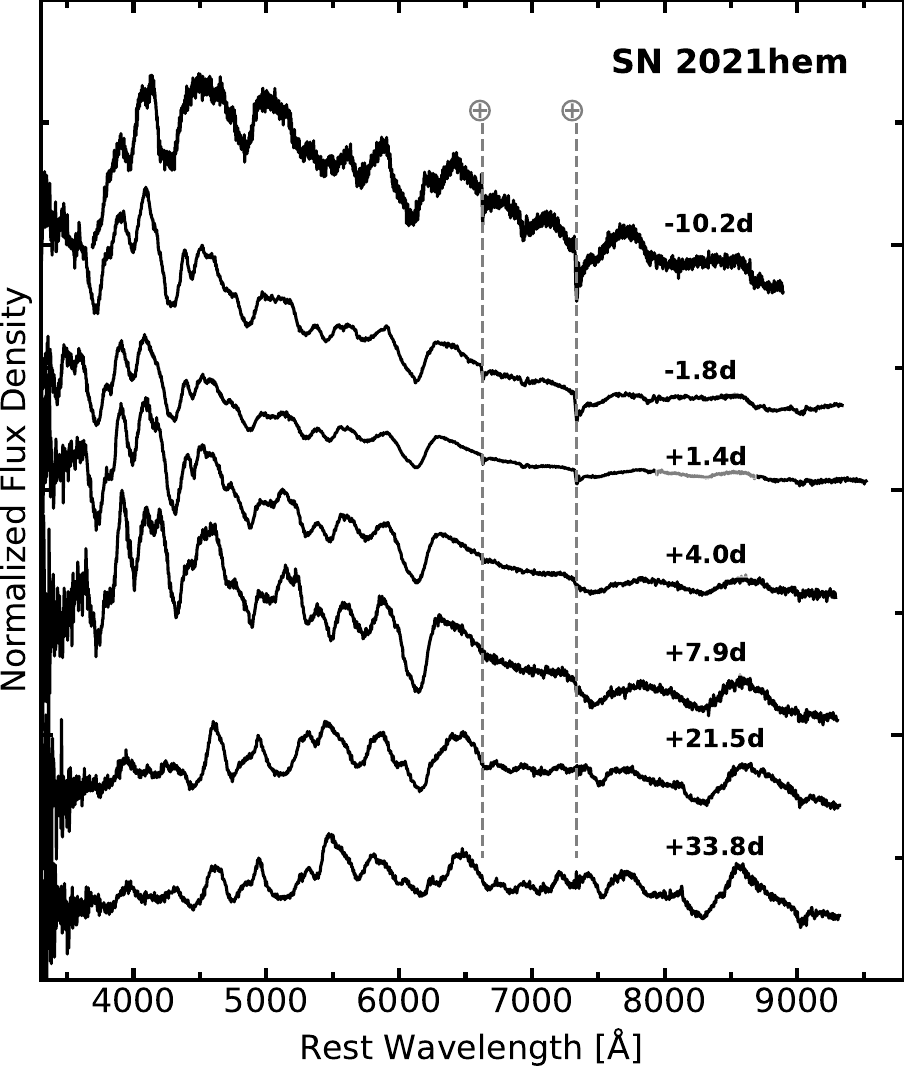}
\caption{Optical spectra of \sn\ from $-$10.2~days to $+$33.8~days relative to the epoch of $B$-band maximum. Spectra are shifted by arbitrary constants for clarity. Telluric absorption features are marked by dashed lines and labeled with  $\oplus$ symbols.} 
	\label{fig:optspec}
\end{figure}

\begin{figure}
\includegraphics[width=\linewidth]{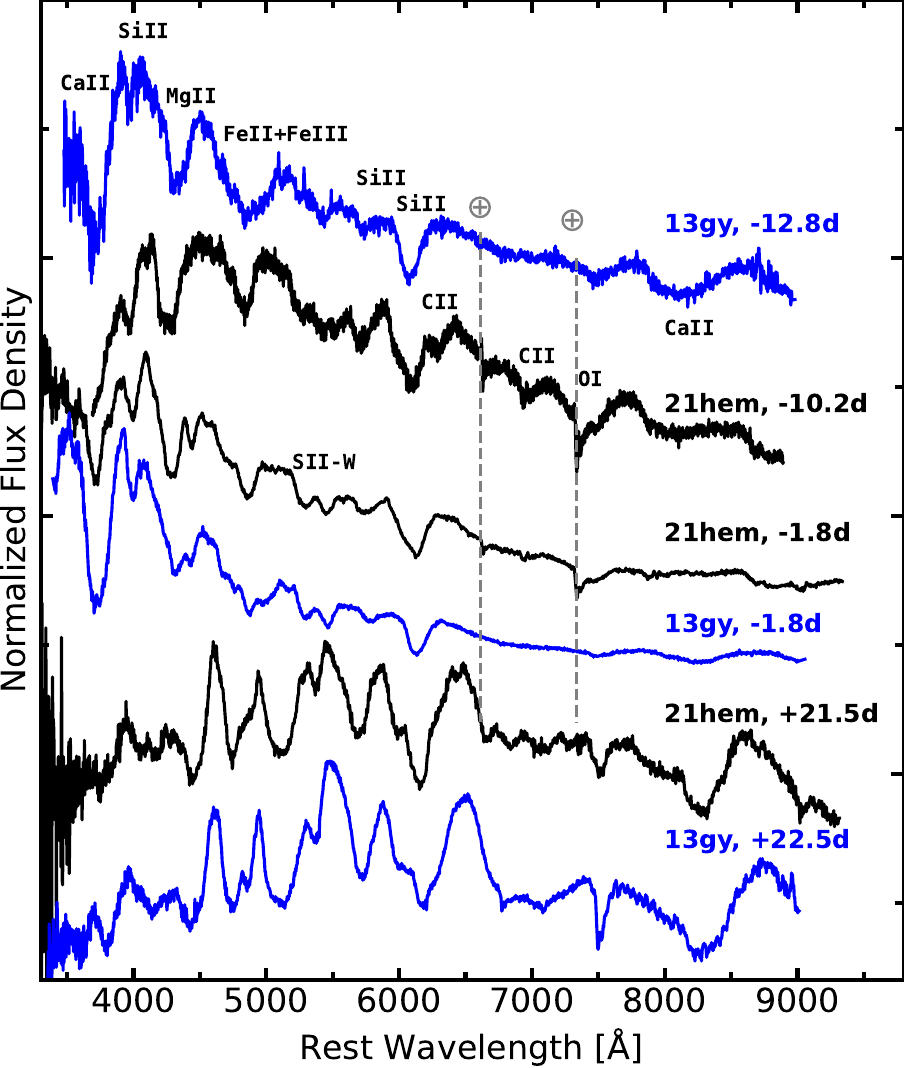}
\caption{Comparison of similar phase spectra of the normal SN~2013gy and \sn. Telluric absorption features are marked by dashed lines and labeled with  telluric symbols. Prominent spectral features in the pre-maximum light spectra are labeled.}  
\label{fig:speccomparison}
\end{figure}

\subsection{Spectral properties, line diagnostics and comparative analysis}
\label{sec:specanal}

Fig.~\ref{fig:optspec} presents our montage of seven optical spectra of \sn, spanning from $-10.2$ to $+33.8$ days. The spectra reveal characteristic SNe~Ia features with the shape of the continuum evolving from the blue to the red over time as the time-series tracks the temperature evolution of the underlying emission regions. Close inspection of the $-10.2$~days spectrum reveals a prevalent \ion{C}{ii} \ld6580 feature appearing in the red wing of the most prominent feature \ion{Si}{ii} $\lambda6355$ (see line identifications in Fig.~\ref{fig:speccomparison}). Such a feature is estimated to occur in approximately 30\% of SNe~Ia \citep{Parrent2011,Folatelli2012,Silverman2012,Maguire2014} and is typically associated with the presence of unburned carbon. However, it appears to be ubiquitous among 2003fg-like SNe~Ia \citep[e.g.,][]{2007ApJ...669L..17H,2011MNRAS.412.2735T,Ashall2021}, which tend to have carbon-rich ejecta, and in some cases, it persists even beyond the time of maximum light (see below for more details).
The measured pseudo-equivalent width ($pEW$) of the \ion{C}{ii} feature from the  $-10.2$\,days spectrum  is $\sim8.9\pm2.5$\,\AA. This is about a factor of 4 higher than that inferred from the normal SN~2011fe, and more consistent with values inferred from similar epoch spectra of the 2003fg-like SNe~2009dc and 2012dn \citep[see][]{Ashall2021,Lu2021}. The measured   velocity of \ion{C}{ii} is $-v_{abs} = 13800\pm560$\kms\ is also in agreement with measurements inferred from similar epoch spectra of the 2003fg-like objects including SN~2009dc, ASASSN-15hy, and ASASSN-15pz \citep{Ashall2021,Lu2021}. We also identify the feature near 7000\, \AA\ likely to be \ion{C}{ii} \ld7235 line, %
which is roughly at a velocity consistent with the \ion{C}{ii} \ld6580 line.
Beyond the \ion{C}{ii} features, the line features present in the  time-series of \sn\ are  consistent with  many other normal SNe~Ia.

Figure~\ref{fig:speccomparison} shows a comparison of \sn\ spectra at representative phases with those of the spectroscopically normal SN~2013gy \citep{Holmbo2019}. The comparison includes \sn’s earliest spectrum at $-10.2$~days, one near maximum light ($-1.8$ days), and another obtained approximately three weeks post-maximum. Key spectral features in the early-time spectra are labeled with their associated ions to aid in identification. Overall, \sn\ closely resembles SN~2013gy at all three epochs. The most notable difference is the presence of \ion{C}{ii} absorption in the pre-maximum spectra of \sn, which is absent in SN~2013gy.

We now examine the standard spectroscopic diagnostics used for  SNe~Ia to determine where \sn\ fits within the established subtypes. This includes measuring the pEWs of the \ion{Si}{ii} \ld5972 and \ld6355 features, which enables classification according to the \citet{Branch2006} scheme. We also measure the Doppler expansion velocity of the \ion{Si}{ii} $\lambda6355$ feature near maximum light, which provides insight into the ejecta kinematics. 

The Branch diagram provides a quantitative framework for spectroscopic classification, relying on the pEWs of the \ion{Si}{ii} $\lambda5972$ and $\lambda6355$ absorption features. Within this parameter space, SNe~Ia are distributed into four principal subgroups: Core Normal (CN), Broad Line (BL), Cool (CL), and Shallow Silicon (SS), which reflect systematic variations in line strength and effective photospheric temperature \citep{Nugent1995}. The pEWs values of the \ion{Si}{ii}  $\lambda5972$ and $\lambda6355$ were measured from the $-1.8$ days spectrum by performing Gaussian fits, yielding values of $21.8 \pm 1.8$\,\AA\ and $111.2 \pm 1.0$\,\AA, respectively. These values place \sn\ firmly within the Branch diagram’s ``core normal'' subclass. 2003fg-like SNe are mostly associated with ``shallow silicon'' group in the branch diagram, while a few, for example, SNe 2012dn and CSS140501, are found to be associated with ``core normal'' group \citep{Ashall2021}. However, \sn\ has \ion{Si}{ii} pEWs larger than any of these SNe, further highlighting the diversity among 2003fg-like SNe.

Using the \texttt{SNIaDCA}\footnote{https://github.com/anthonyburrow/SNIaDCA} code from \citet{Burrow2020}, we find that \sn\ has an $>85\%$ likelihood of belonging to the ``core-normal'' group. When additional components (i.e., $v_{\rm Si~II}$ and $M_B$) are included in the clustering analysis, the probability of \sn\ belonging to the ``cool''-equivalent group increases dramatically. See Appendix \ref{sec:branch} for further details on the probabilistic placement of \sn\ within these multi-dimensional clusters.

\begin{figure}
\includegraphics[width=1\linewidth]{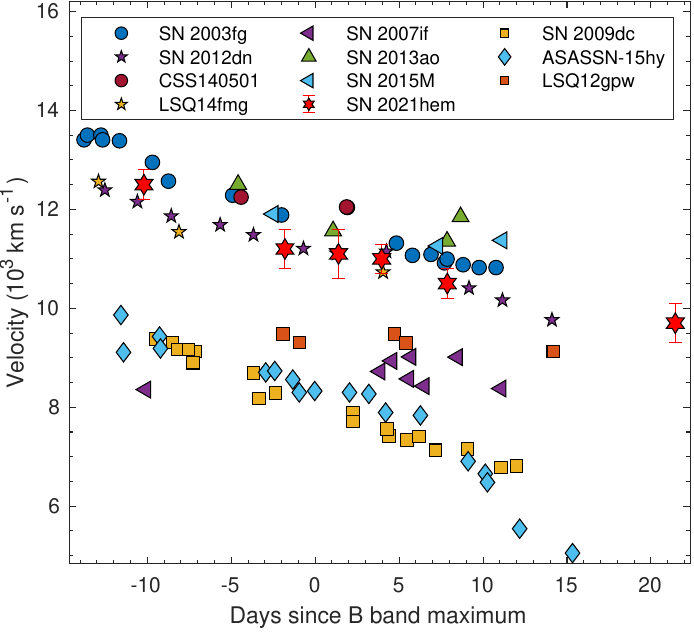}
\caption{\Siii\ \ld6355 velocity evolution of \sn\ compared with other 2003fg-like SNe~
Ia.}
\label{fig:velocity_comp}
\end{figure}

Following the \citet{Wang2009} framework, SNe~Ia can be categorized as high-velocity (HV), with a blueshifted Doppler line velocities inferred from the position of maximum absorption (hereafter  $-v_{abs}$) of the \ion{Si}{ii} line with $-v_{abs}$ $\gtrsim 11{,}800$~km~s$^{-1}$, or  otherwise ``normal'', which populate a narrow strip in the \ion{Si}{ii} velocity distribution, with an average  $-v_{abs} \sim 10{,}600\pm400$~km~s$^{-1}$. %
Turning to  \sn,   we infer from the \ion{Si}{ii} absorption component in the  $-10.2$ days spectrum 
$-v_{abs} \sim 12{,}500 \pm 300$ km~s$^{-1}$. This decreases to  $11{,}200 \pm 400$~km~s$^{-1}$ by $-1.8$ days, and further declines  to  $9{,}200 \pm 300$~km~s$^{-1}$ by +33.8 days. The velocity achieved around maximum is consistent with \sn\ being a normal SN~Ia within the classification framework of \citet{Wang2009}. Figure~\ref{fig:velocity_comp} compares the \ion{Si}{ii} \ld6355 velocity evolution of \sn\ with a sample of 2003fg-like SNe from \citet{Ashall2021}. The velocity evolution is consistent with other 2003fg-like SNe, and particularly identical to SN~2012dn.

\begin{figure}
\includegraphics[width=1\linewidth]{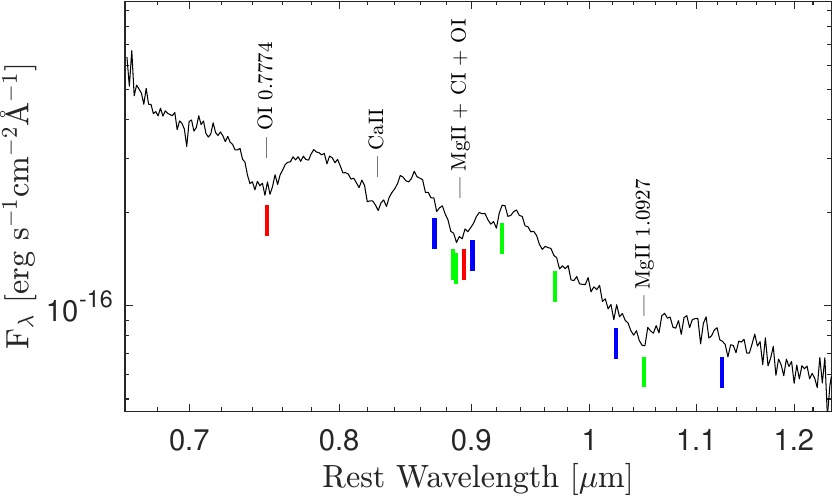}
\caption{The -0.5\,d NIR spectrum of SN~2021hem obtained with the NASA IRFT telescope equipped with SpeX. The prominent identified lines are labeled. The vertical red, green, and blue lines mark the positions for \ion{O}{i}, \ion{Mg}{ii}, and \ion{C}{I} lines respectively, where all lines for a given atomic species are at same velocity. The \ion{O}{i} (red) \ld0.7774, \ld0.9266 \mum\ lines are at 10,500\kms, \ion{Mg}{ii} (green) \ldld0.9218,0.9244, \ld0.9632, \ld1.0092, \ld1.0927 \mum\ lines are at 11,700\kms, and \ion{C}{i} (blue) \ld0.9093, \ld0.9406, \ld1.0693, \ld1.1754 \mum\ lines are at 12,500\kms. Strong lines of \ion{Mg}{ii} and \ion{O}{i} are identified, whereas for \ion{C}{i} no prominent line identification is found, the \ion{C}{i} line positions are shown at a tentative velocity of $\sim12,500\kms$. }
\label{fig:nirspectrum}
\end{figure}

Figure~\ref{fig:nirspectrum} presents the single NIR spectrum obtained at $-0.5$~days with the NASA IRTF equipped with SpeX. The most prominent feature, located just redward of 1.0~$\mu$m, is identified as \ion{Mg}{ii} 1.0927$~\mu$m, with a measured $pEW = 53.5 \pm 5.6$~\AA\ and absorption velocity of $-v_{abs} = 11{,}700 \pm 200\kms$. Common lines features of \Oi\ \ld0.7774 \mum, and \Caii\ NIR triplet are present. Another prominent feature near 0.9\mum\ is visible, which is primarily dominated by \ldld0.9218,~0.9244\mum\ blended with \Oi\ \ld0.9266\mum\ and also possibly \Ci\ \ld0.9406\mum. No clear identification of \Ci\ lines was possible in the spectrum, as \Ci\ lines are usually stronger in 2003fg-like SNe at early times, and tend to become weaker near maximum. This is also consistent with the absence of \ion{C}{ii} in optical spectra of \sn\ near maximum light. However, in Fig.~\ref{fig:nirspectrum}, line positions of \Ci\ lines are marked at a tentative velocity of $-v_{abs} = 12{,}500\kms$. Some weak features may be associated with the \Ci\ line positions in the figure, however, the SNR level of the spectrum limits any conclusive identification.  

In summary, the spectroscopic line diagnostics of \sn\ are aligned with spectroscopically normal SNe~Ia, except for the relatively prevalent \ion{C}{ii} line present in our earliest optical spectrum, which is commonly seen in 2003fg-like SNe. 
The conclusion drawn from our near maximum-light spectra of \sn\ is consistent with many 2003fg-like SNe, as at this phase often these objects are spectroscopically indistinguishable from normal SNe~Ia \citep[see Sect.~6 of][]{Ashall2021}.

\subsection{Revisiting redshift measurement and potential biases}
\label{sec:redshiftbias}

It is important to note that all absorption-minimum velocities quoted in this section include an additional systematic uncertainty of $\Delta v_{\rm sys} = \pm 1{,}500~\kms$, propagated from the uncertainty in the redshift adopted in Sect.~\ref{sec:redshift}. 2003fg-like SNe exhibit a range of \Siii\ velocities, from relatively low values of $\sim8{,}000~\kms$ \citep[e.g.][]{2006Natur.443..308H,2009ApJ...707L.118Y} to more typical SNe~Ia-like values up to $\sim12{,}000~\kms$ \citep[e.g.][]{2007ApJ...669L..17H}. Since the \sn\ redshift was estimated by matching SN~Ia spectral templates using SNID (see Sect.~\ref{sec:redshift}), an intrinsically low-velocity SN could be matched to a normal-velocity template, thereby biasing the inferred redshift toward higher values.
This, in turn, could lead to an overestimation of the absorption-minimum velocities from spectra corrected to rest-wavelength using the biased redshift. On the other hand, no significant bias is expected in template matching if the SN is intrinsically a normal velocity SN, which is on the higher end of 2003fg-like SNe velocities. 
Based on the analysis done so far, \sn\ appear to be an SN in the normal velocity range, yet, we are unsure if the measurements are biased by overestimated redshifts. 

To investigate this potential bias, we examine the \Siii\ $\lambda6355$ absorption line FWHM. The  FWHM velocity ($v_{\rm FWHM}$) can be used as a tracer of the line velocity and is almost independent of the redshift correction applied to the spectrum. This may provide a clue whether \sn\ is intrinsically a normal-velocity or low-velocity 2003fg-like SN. Although the $v_{\rm FWHM}$ may not be mapped directly to absorption-minima velocity, it can serve as a proxy to relate with absorption-minima velocity. We measure both the \Siii\ \ld6355 $-v_{\rm abs}$ and $v_{\rm FWHM}$ for a number of 2003fg-like SNe spanning a range of velocities. As until the maximum light \Siii\ \ld6355 line might be contaminated by the \ion{C}{ii} line, we use coeval post-maximum spectra for this diagnostic.
In the case of SN~2009dc at +4.5~days, $-v_{\rm abs} = 7{,}300 \pm 150~\kms$ and $v_{\rm FWHM} = 5{,}300 \pm 200~\kms$; for ASASSN-15hy at +4.2~days, $-v_{\rm abs} = 7{,}900 \pm 150~\kms$ and $v_{\rm FWHM} = 5{,}900 \pm 200~\kms$; for LSQ14fmg at +3.6~days, $-v_{\rm abs} = 10{,}400 \pm 200~\kms$ and $v_{\rm FWHM} = 6{,}300 \pm 250~\kms$; and for SN~2012dn at +3.9~days, $-v_{\rm abs} = 10{,}500 \pm 200~\kms$ and $v_{\rm FWHM} = 6{,}400 \pm 250~\kms$. These values show a monotonic relation between $-v_{\rm abs}$ and $v_{\rm FWHM}$.
Now turning to \sn\, the estimated velocities for +3.9 days spectrum are $-v_{abs}=11,000\pm350\kms$ and $v_{\rm FWHM}=8{,}600\pm250\kms$. 
In the case of \sn, the measured $v_{\rm FWHM}$ is consistent with that of normal and higher-velocity 2003fg-like SNe. In fact, $v_{\rm FWHM}$ of \sn\ exceeds that of any of the measured 2003fg-like SNe, likely due to its extended blue-wing in the absorption line profiles, a feature not observed in any of the comparison objects. However, this analysis indicates that \sn\ is not an intrinsically low-velocity 2003fg-like SN, and therefore it is unlikely that the \texttt{SNID}-based redshift estimate is affected by a significant systematic overestimate.

\section{Search for the host galaxy}
\label{sec:hostgalaxy}
One of the most intriguing aspects of \sn\ is its apparently hostless environment. SNe~Ia are known to occur in a wide variety of galaxies and environments \citep{2006ApJ...648..868S,Lampeitl2010,Pan2014}, including stellar populations within galaxy clusters, which have been stripped from their hosts through tidal interactions and/or merger events. Only a handful of such intracluster SNe~Ia have been identified to date \citep{2003AJ....125.1087G, 2011ApJ...729..142S, 2015ApJ...807...83G}. In the case of \sn, no nearby galaxy cluster is present in the field, making its origin from an intracluster star unlikely, although this possibility cannot be excluded. 
\citet{2025ApJ...988..278S} investigated the likely origins of SNe~Ia that appear to lack associated host galaxies, particularly those occurring in open-field, non-clustered environments, and suggested that many of these seemingly hostless events are actually associated with faint dwarf galaxies that remain undetected in wide-field surveys due to shallow limited imaging depth.

This section examines whether \sn\ originated from a hyper-velocity progenitor that traveled far from its parent galaxy, or alternatively from an undetected low-luminosity host galaxy.

\subsection{A runaway progenitor from its host}
\label{sec:arunawayburnoutnooneloved}

Unlike core-collapse SNe, SNe~Ia exhibit long explosion delay times, typically ranging from $\sim 100$~Myr to several Gyr, with a delay-time distribution that follows approximately $t^{-1}$ \citep{2008PASJ...60.1327T,2012PASA...29..447M}. In principle, such timescales allow a hyper-velocity progenitor to travel several hundred kiloparsecs from its host galaxy prior to explosion. To date, however, no confirmed case of a hyper-velocity progenitor has been identified.

\bose{Assuming \sn\ originated from a hyper-velocity runaway star (or binary system), we searched for known or visible galaxies in the vicinity of \sn\ as a plausible host. 
Two of the galaxies have been identified from SIMBAD \citep{2000A&AS..143....9W} catalog as the closest and equidistant candidates from \sn\ (see Fig.~\ref{fig:findingchart}). One is an elliptical galaxy LEDA~1461683 ($\alpha = 16^h21^m25^s.051$; $\delta = +14^\circ34'45\farcs60$) at $z = 0.02873$, and a projected distance of $102\kpc$. The other is a LINER-type Active Galaxy Nucleus (AGN), 2MASX~J16212572{+}1431537 ($\alpha = 16^h21^m25^s.723$; $\delta = +14^\circ31'53\farcs73$), which is at a similar projected distance of 104\kpc\ and redshift of $z = 0.02952$.
Although both of these galaxy redshifts differ from the plausible galaxy group's redshift of $z\sim 0.035$ or the adopted SN redshift of 0.0363 as inferred in Sect.~\ref{sec:redshift}, such a discrepancy could be explained if the progenitor were a hyper-velocity star. A star with a line-of-sight velocity of $\sim 2000$~\kms\ would account for the higher observed redshift. As an interesting side note, considering the redshift and projected line-of-sight velocity of the star in this scenario, after accounting for reddening, luminosity distance, and K-correction,
the absolute rest-frame magnitude of  \sn\ would be $M_B=-19.45$\,mag.
}

\bose{Based on their redshifts and projected separation from the SN position, both galaxies are, in principle, equally plausible hosts of \sn. However, considerations of galaxy type and morphology suggest that the AGN host is significantly more likely to produce a hyper-velocity star. The size of the AGN galaxy is roughly $\sim$2.3 times that of the other elliptical galaxy, implying a larger stellar halo and also a smaller normalized directional light distance to \sn. Thus making the AGN galaxy the higher probability candidate among the two, to be the associated host of \sn. Although AGN jets or winds cannot directly accelerate compact, massive objects such as stars to extreme velocities, AGNs can facilitate the production of hyper-velocity stars with speeds up to $\sim$4,000\kms\ via the \citet{1988Natur.331..687H} mechanism, in which a binary star system is disrupted by a supermassive black hole (SMBH) ejecting one of the members at high velocity. Additional channels are also possible, including star formation within the medium of AGN-driven outflows \citep{2018NewA...61...95W}, or stellar ejection by interaction with binary SMBHs \citep{2003ApJ...599.1129Y, 2019MNRAS.482.2132D}.
Furthermore, AGN hosts typically exhibit significantly elevated star-formation rates, particularly in the central region \citep{2025MNRAS.539.3229G}, which would naturally increase the number of stars available for dynamical interactions with the SMBH and thereby enhancing the expected production rate of hyper-velocity stars. 
Considering these factors collectively, the AGN galaxy 2MASX~J16212572{+}1431537 appears to be the most plausible origin of the hyper-velocity progenitor (if such a scenario is assumed) of \sn. %
}

\bose{Assuming a delay time of $\gtrsim$100\,Myr and the progenitor traveled a projected distance of 104\kpc\ from the AGN host, the projected transverse velocity is $\lesssim$1{,}020\kms. This, combined with the line-of-sight velocity component of $2{,}000\kms$ (as discussed above), corresponds to an un-projected 3-D velocity of $\lesssim 2{,}240 \kms$.  In comparison, the fastest known Galactic hyper-velocity star is 1{,}800\kms\ \citep{Koposov2020}. A larger delay time can reduce the expected velocity of the hyper-velocity progenitor. However, the adopted delay time of $\sim$100\,Myr is at the lower limit, but is also the highest probable value as the delay time distribution roughly follows $t^{-1}$. A lower redshift than that inferred for the SN would also lessen the required line-of-sight velocity component and, consequently, the net 3D velocity. Thus, assuming the minimum of $\sim$100\,Myr delay time for \sn\ explosion, the required maximum velocity of the hyper-velocity star is $\sim 2{,}240 \kms$, which is only $\sim$400\kms\ higher than the fastest known Galactic hyper-velocity star, and is within the theoretical limit of fastest ejected stars by an AGN host.
}

\begin{figure}
\includegraphics[width=1\linewidth]{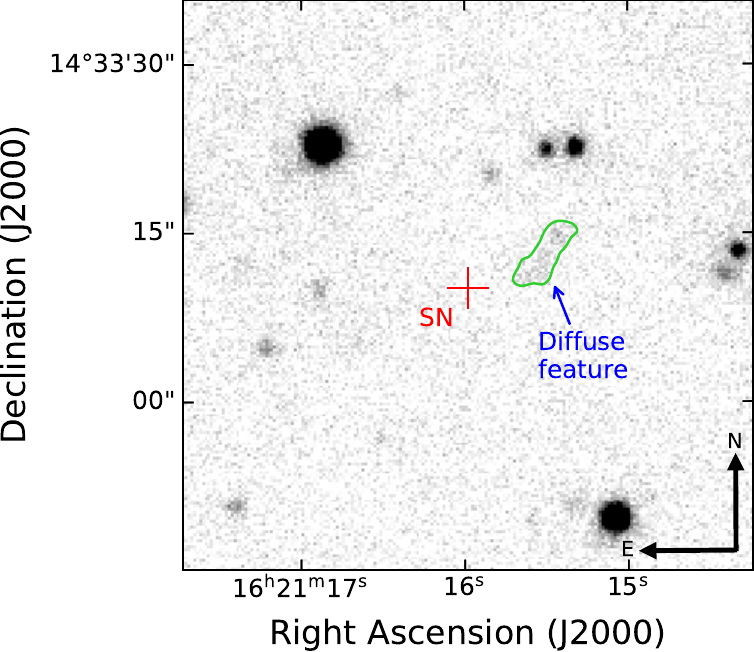}
\caption{Image cutout centered on the position of \sn, taken on 9 September 2023 with the GTC equipped with  OSIRIS. The final image is a stack of eight individual exposures, with a total integration time of 435 seconds. The location of the SN and the nearby diffuse extended feature is marked. An isophote is shown (green contour) enclosing the diffuse feature.}
\label{fig:GTCimage}
\end{figure}

\subsection{An undetectable faint dwarf host}

We also explored the possibility that the SN resides in a low-luminosity dwarf host too dim to be detected in archival data. However, neither image analysis nor visual inspection revealed any trace of a host galaxy at the SN location, even in our deepest exposures. This makes it important to establish an upper limit on the host-galaxy brightness below which a source can not be detected in our data.

\subsubsection{Determining the limiting magnitude from the GTC image}
\label{sec:limmag}

We use the 435s $r$-band GTC image shown in Fig.~\ref{fig:GTCimage}, obtained approximately 900 days after the explosion, to determine the limiting magnitude. After processing the image following the standard reduction procedure, image quality parameters: stellar FWHM and ellipticity are measured with SExtractor, and median sigma clipping is applied to reject nonstellar sources. Aperture photometry is then performed using the photutils DaoPhot package. The image is calibrated using field stars cross-matched with the PanSTARRS $r$-band catalog \citep{Flewelling2020}, and the zero-point magnitude for the image is determined to an uncertainty of $0.01$\,mag. 

To determine the limiting magnitude, an artificial test star of a given apparent magnitude is inserted near the SN location. Detectability is evaluated through $3\sigma$-threshold source detection with SExtractor \citep{1996A&AS..117..393B} within 0.5$\cdot$FWHM radius, followed by aperture photometry to assess recovered flux accuracy. A reference artificial star, fixed at 19 mag and placed in a region free of nearby field stars, is used to correct for flux loss due to fixed-aperture photometry. The procedure is repeated for 30,000 realizations, with test star apparent magnitudes drawn uniformly between 21.75 and 24.50 mag and positions randomized within a $4\cdot$FWHM width ($\sim4\farcs7$) square centered on the SN location. To improve computational efficiency and reduce the impact of large-scale background artifacts, a 500\,pix cutout centered around the SN location is used.

\begin{figure*}[!t]
\centering
\includegraphics[width=0.494\textwidth]{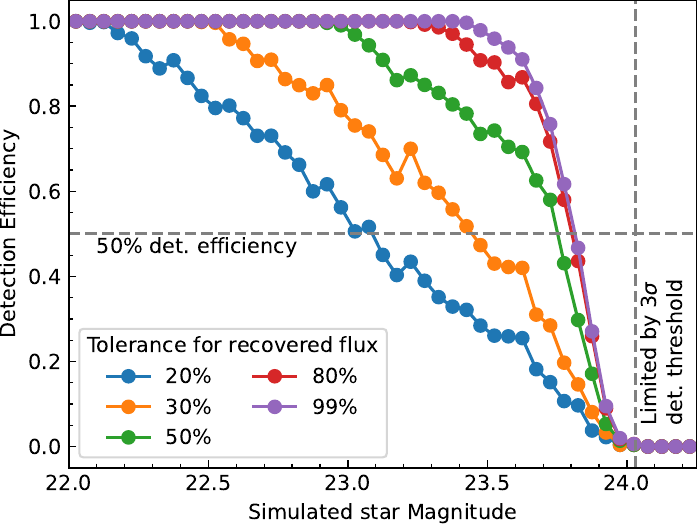}
\hspace{0.003\textwidth}
\includegraphics[width=0.494\textwidth]{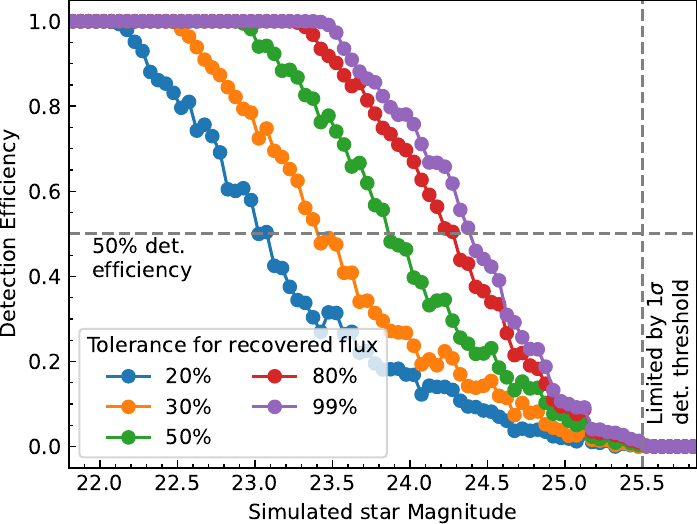}
\caption{\textit{Left:} Detection efficiency as a function of magnitude with $3\sigma$ detection threshold. Each curve is based on 30,000 simulated images, generated by adding an artificial star drawn from a uniform distribution over the magnitude range and within a $\rm 4 \cdot FWHM$ square region around the SN position. The different curves correspond to varying tolerances between the recovered flux and the true flux of the input star. A detection is counted as successful if the recovered flux lies within the specified tolerance level of the input artificial star. \textit{Right:} Same as left panel, but using $1\sigma$ detection threshold and 50,000 simulations within magnitude range of 21.75 to 25.75\,mag.}
\label{fig:deteff}
\end{figure*}

The detection efficiency curves are plotted in the left panel of Fig.~\ref{fig:deteff}, where each curve corresponds to different tolerances of recovered flux error with respect to the input artificial test star's flux. For example, a 20\% tolerance of recovered flux implies that a detection is considered true only if recovered flux is within 20\% tolerance of its true flux of the artificial star. Similarly, the 99\% tolerance curve implies the efficiency curve is almost solely determined if stars are detected above $3\sigma$ noise threshold. At the cutoff magnitude of $m_{cutoff,r}\approx24.05$\,mag, detection efficiency becomes zero for all the curves, as the test star is too faint to cross $3\sigma$ noise threshold at any location within $\sim 4\farcs7$ square area (as mentioned above).

Since we are primarily interested in detectability of a source (host galaxy) near the SN location, irrespective of the accuracy of the measured flux, we look into the 99\% tolerance curve at 50\% efficiency to determine the limiting-magnitude, and thus a value of  $m_{lim,r}=23.8$\,mag. This implies, assuming a point-source, any host associated with the SN is fainter than 23.8\,mag, which translates to an absolute magnitude of $-12.3$\,mag by adopting a luminosity distance for redshift of 0.0363.

It is important to emphasize that, in addition to image quality, the limiting magnitude we adopt also depends on the constraints imposed in our photometric workflow. The detection threshold is one of the primary parameters that governs the detectability of point sources. 
A second consideration is the definition of “sufficiently accurate” photometry, which we have already addressed through the generation of multiple efficiency curves for different tolerance levels (see Fig.~\ref{fig:deteff}), where each curve represents a different tolerance in accuracy of recovered flux. In most cases, a $3\sigma$ detection threshold provides a robust balance between accuracy and efficient detection, and this criterion was applied in the preceding analysis.

However, in this work, our primary interest lies in establishing the detectability of a possible host galaxy rather than obtaining reliable photometry of its features. For this purpose, a lower threshold is justified, and we find that a $1\sigma$ detection threshold represents a reasonable limit. For example, the faint diffuse feature highlighted in Fig.~\ref{fig:GTCimage} (see Sect.~\ref{sec:diffused_feature}) lies well below the $3\sigma$ noise level, but is detectable at $1\sigma$ level and also by visual inspection. 

We therefore repeated the limiting-magnitude analysis using a $1\sigma$ detection threshold, with the results shown in the right panel of Fig.~\ref{fig:deteff}. The notable difference with a lower detection threshold is the fainter cutoff magnitude at $m_{cutoff,r}\approx25.50$\,mag, and the efficiency curves decline more gradually towards faint magnitudes. From the analysis with $1\sigma$ detection threshold, at 50\% efficiency for the 99\% tolerance curve, the limiting magnitude we find is $m_{lim,r}=24.4$\,mag, or an absolute magnitude limit of $-11.6$\,mag for any undetected galaxy.

\subsubsection{Analytical estimate of limiting magnitude}
As a sanity check for the limiting magnitude estimated above, we attempt to analytically calculate the limiting magnitude for the image based on the detection threshold. Assuming a Gaussian stellar PSF, the 2D profile is
\begin{equation}
	f(x,y) = h \cdot exp{\left[ -\frac{(x - \bar{x})^2}{2\sigma_x^2} - \frac{(y - \bar{y})^2}{2\sigma_y^2}\right]}, 
\end{equation}
where $h$ is the amplitude of the Gaussian profile, and other symbols have their usual meanings. The volume of the 2D profile would be,
\begin{equation}
	V= h\cdot 2\pi \sigma_x \sigma_y.
\end{equation}

\noindent For a circular symmetric stellar PSF, $\sigma_x = \sigma_y = \sigma_{PSF}$. To satisfy a $3\sigma$ detection threshold, the amplitude $h$ must be $> 3\sigma_{sky}$, where $\sigma_{sky}$ is the local background sky-noise. Therefore, the limiting counts in ADU for the profile to satisfy $3\sigma$ detection threshold,
\begin{equation}\
	C_{lim} = 6\pi\sigma_{sky}\sigma_{PSF}^2, 
\end{equation}
and on converting to limiting magnitude, 
\begin{equation}
	m_{lim} = ZP - 2.5log_{10}\left[  6\pi\sigma_{sky}\sigma_{PSF}^2 \right],
\end{equation}
where $ZP$ is the image zero-point magnitude for one ADU. Upon substituting the measured parameters for GTC image,  $\sigma_{sky}=27.12$\,ADU estimated from a 3$\cdot$FWHM circular radius at the location of the SN, the $\sigma_{PSF} = 1.95$\,pix measured for stellar PSF of the image, and the $ZP= 32.61$\,mag, we estimate a limiting-magnitude, $m_{lim}=24.38$\,mag. This is roughly consistent with the limiting magnitude cutoff obtained from simulation for the $3\sigma$ detection threshold limit, which is $m_{cutoff}\approx24.05$\,mag.
In the case of a  $1\sigma$ detection threshold, the corresponding value is $m_{lim}=25.57$\,mag, which is very close to the magnitude cutoff of $m_{cutoff}\approx25.50$\,mag determined from $1\sigma$ threshold simulation.

\subsubsection{Limiting magnitude in terms of surface brightness}
\label{sec:lim-sb-mag}

In the case of extended objects like galaxies, the detection limit can be better described in terms of surface brightness. Assuming a circular aperture area $A_{ap}$ within which the flux for the estimated point-source limiting magnitude is confined, the limiting surface-brightness can be approximated to be fainter by $\rm 2.5 log_{10}(A_{ap})\sbunit$. 

This limiting surface brightness should be regarded as an upper limit for the image.  On considering a source that is extended over a larger area on the sky, the total integrated flux increases, while the noise per arcsec$^2$ decreases by a factor of $1/\sqrt{Area}$ due to averaging. Consequently, for extended objects such as galaxies, and also depending on their morphology, it becomes possible to detect objects with fainter surface brightness. For surface brightness measured for a extended source over an area of $A_{sb}$, the S/N would scale up as $\sqrt{A_{sb}/A_{ap}}$, and the limiting surface-brightness would be fainter by a additional factor of  $1.25 \log_{10}(A_{sb}/A_{ap})\sbunit$. Therefore, the limiting surface-brightness  would be:

\begin{equation}
	\mu_{lim}= m_{lim} + 2.5 \log_{10}(A_{ap}) + 1.25 \log_{10}\left(\frac{A_{sb}}{A_{ap}}\right)\sbunit,
\end{equation}
\noindent or
\begin{align}
\mu_{lim} &= m_{lim} + 2.5 \log_{10}\!\left(\tfrac{\pi}{4}\,{\rm FWHM}^2\right) \nonumber \\
          &\quad + 2.5 \log_{10}\!\left(\frac{D_{sb}}{{\rm FWHM}}\right)\,\sbunit ,
\end{align}
\noindent or
\begin{equation}
	\mu_{lim}= m_{lim} + 2.5 \log_{10}(\pi/4~{\rm FWHM}\cdot D_{sb})\sbunit,
\end{equation}

\noindent where FWHM is the point-source FWHM in arcsec, and $D_{sb}$ is the diameter of the extended source in arcsec over which the surface-brightness is to be estimated. For our image, the equation reduces to $\mu_{lim}= m_{lim} + 2.5\log_{10}(D_{sb})-0.094$. 

Assuming 5\kpc\ effective diameter for a UDG-like host, the corresponding angular size at the redshift of the SN is 6\farcs2, and the limiting surface-brightness for the image is fainter by 1.9\sbunit\ than the point source limiting magnitude. Adopting the $1\sigma$ point source limiting magnitude determined in Sect.~\ref{sec:limmag}, the limiting surface-brightness for a 5\kpc\ extended source near the SN location in the image is $\mu_{lim,r}=26.3\sbunit$.

There are a few caveats of this limiting surface-brightness estimate that should be noted. For a real galaxy, the limit of the surface brightness measurement would also be affected by the small-scale surface brightness variation of the galaxy. However, UDGs or UFDGs do not show much morphological diversity, and hence would not have a significant effect on our conversion.  Secondly, when converting from a point source limiting magnitude to a  surface-brightness limiting magnitude, we assume the sky background properties of the extended source, which contribute to detectability, are similar to those of the point source aperture and the sky annulus background. This also does not take into consideration the possible large variation in the sky background when an extended source area is being considered. However, this issue is somewhat mitigated, as in our point-source limiting magnitude simulation, we used a square area of $\sim 4\farcs7$ (3.8\kpc) width.

The above estimated limiting surface-brightness magnitude implies that a galaxy of diameter 5\kpc\ (or equivalent area) with 26.3\sbunit\ surface-brightness is detectable to $1\sigma$ limit. Any galaxy of larger angular size can be detected even to a fainter surface brightness limit, and vice versa.  
In this calculation, we adopted the diameter of the extended source to be $\sim5$~kpc, which is roughly the median effective diameter found for UDGs \citep[see e.g.,][]{2018MNRAS.473.3747S}. Having this limiting surface-brightness magnitude, we can rule out the possibility for almost all undetected UDGs as a host galaxy, except for the ones that are at the faintest end of UDG luminosity function $26.3 \lesssim \mu\lesssim 26.5\sbunit$. However, we can not rule out the possibility of a UFDG-like host, that might have $\mu>27\sbunit$. Furthermore, UFDGs may have radii only up to a few hundred parsecs \citep{2019ARA&A..57..375S}, which would be a point source at our image resolution, and thus limited by point-source limiting magnitude of $m_{lim,r}= 24.4$\,mag.

\subsection{A nearby faint diffuse feature}
\label{sec:diffused_feature}

A faint, diffuse, and extended feature is detected in the deep GTC image at a projected distance of 5\farcs3 ($\sim4.2$\,kpc) from the SN location, measured from its nearest edge (see Fig.~\ref{fig:GTCimage}). This feature is not visible in any of the archival images, nor in our shallow follow-up imaging. The feature extends 6\farcs3 ($\sim5.0$\,kpc) along its longest axis. Although its nature and possible association with the SN remain uncertain, photometry  of the diffuse feature was performed using a polygonal region enclosing the entire structure (see Fig.~\ref{fig:gal_phot_reg}). For background estimation, a similarly shaped polygonal annulus of width $1.2 \times$ FWHM was defined around the feature. Calibration was carried out using the zeropoint determined in the previous section. The integrated apparent magnitude of the feature is estimated as $m_r=22.8 \pm 1.8$\,mag, corresponding to an absolute magnitude of $M_r=-13.2 \pm 1.8$\,mag. Furthermore, the surface brightness of the feature was estimated by measuring its brightest regions (highlighted in red in Fig.~\ref{fig:gal_phot_reg}), yielding a value of $\mu_r=25.58 \pm 0.70$\sbunit. 
This inferred value falls within the luminosity distribution of UDGs \citep[e.g.,][]{vanDokkum2015,Koda2015,2017A&A...607A..79V,Newton2023}. Together, these comparisons suggest that if the feature is indeed a galaxy, it would be broadly consistent with UDGs found residing within clusters.  
However, the feature appears unusually elongated and irregular in shape with its projected axis ratio of $0.25 \pm 0.05$. Typically, UDGs are not thin or flat disks, their shapes tend toward thick, spheroidal, or elongated spheroids rather than thin oblate disks \citep[e.g.,][]{2017ApJ...851...27M,2020ApJ...899...78R}. Only a small number of UDGs, those in galaxy clusters, show distorted, elongated, or lopsided shapes due to tidal disruptions with a nearby massive galaxy \citep[e.g.,][]{2021A&A...654A.105M}.
In comparison, typical UDGs exhibit axis ratios in the range $0.4$–$0.9$ with a median value of $\sim0.7$ \citep{2017ApJ...838...93B,2020ApJ...899...78R}. Therefore, the extreme elongation of the diffuse feature near \sn\ is unlikely for an UDG, and could potentially be only explained by tidal interactions with a nearby galaxy, yet no such companion is present in the immediate vicinity of the diffuse feature.

If we assume the feature is a galaxy, the projected distance of SN at $\sim5.5-6.4\kpc$ from the center of the feature (or $\sim4.2\kpc$ from the edge) is unusually high for it to be the associated host, when the small size of this galaxy, or if a typical UGD, is considered. 
We look into the normalized directional light distance \ddlr\ of the SN \citep{2006ApJ...648..868S,2016AJ....152..154G}, which is a measure of the distance between the SN and the host galaxy, normalized to the physical extent of the host galaxy (directional radius of host) in the direction of the SN. For \sn\ and this feature, we measure $d_{DLR}$ to be $3.1^{+1.2}_{-0.2}$. The uncertainty attributed to the measurement is primarily due to the irregular shape of the feature and the region of confusion arising in considering the center of the host galaxy for measurement. This measure essentially means, the distance to the SN is roughly three to four times farther than the visible radius of the host galaxy.
We compared this with the \ddlr\ measurements of $\sim1{,}600$ SNe from ZTF DR2 sample \citep{2025A&A...694A..10D}. The sample is cutoff to $z<0.06$, to ensure higher volumetric completeness, and also found the bias in \ddlr\ values towards higher redshift is minimal with this cutoff imposed as compared the full sample. 
The comparison reveals that only $\approx1.8 - 4.0$ percentiles of SNe~Ia in the sample have potential hosts that are further than the \ddlr\ value measured between \sn\ and this diffused feature. It is to be noted that in ZTF DR2 sample, the potential hosts are identified based on the closest (\ddlr) galaxy candidate identified from Pan-STARRS1 and Legacy imaging surveys, and no attempt was made to detect hosts fainter than survey depths. Therefore, it is possible that the true percentile of hosts above the given \ddlr\ value could be even lower than inferred above. 

Considering the large distance (\ddlr) of the SN from the feature, and also the highly elongated and irregular shape of it, makes the diffuse feature a candidate of low probability to be the associated host of \sn.
However, we cannot completely rule out that the diffuse feature is a galaxy, or if it is, its likelihood to be the associated host of \sn. If this is indeed the host of \sn, this would be the first SN detected in a UDG, and the progenitor has to be a high velocity star to reach this far before explosion. UGDs are low metallicity systems, and that favors being a host for 2003fg-like SNe. An alternative possibility could be that the diffuse feature is an unrelated background galaxy at a higher redshift.

\section{Conclusions}
\label{sec:summary}
The Type~Ia \sn\ was discovered within 48 hours of last non-detection in ZTF survey, in an apparently hostless environment.  Spectral template matching yields a redshift of $z = 0.0363 \pm 0.0049$, which is also consistent with an overdensity of similar redshift galaxies within $\sim$1\,Mpc radius. Its peak absolute $B$-band magnitude of $M_{B, max} = -19.96 \pm 0.29$\,mag places \sn\ at the bright end of the SN~Ia luminosity distribution. Unlike normal or luminous 1991T-like SNe~Ia, however, the $i$- and NIR-band light curves show no secondary maximum -- a hallmark of low-luminosity 1991bg-like SNe~Ia and some other peculiar subtypes. The high peak luminosity combined with the lack of a secondary maximum in the $i$- and NIR-band light curves suggests \sn\ is a 2003fg-like SN~Ia candidate.

The light curves of \sn\ exhibit slow evolution, characterized by shape parameters $\dmb = 1.02 \pm 0.02$ mag and $\sbv = 0.94 \pm 0.05$ mag.
With these parameters, \sn\ is $\sim0.3$\,mag brighter than typical SNe~Ia in both the luminosity–width and luminosity–color–stretch relations.
By fitting the radioactive decay model to the tail of the bolometric light curve of \sn, we estimate a high \nickel\ mass of $M_{Ni}=1.00\pm0.09~M_{\odot}$, and a gamma-ray trapping timescale of $t_{0\gamma}=51\pm3$\,day. Such a \nickel\ mass is significantly higher than that of normal SNe~Ia, but consistent with a more massive progenitor as inferred for several other 2003fg-like SNe. The modeling of the time-weighted integral of the luminosity shows that the entire energy budget of the SN can be solely explained by the radioactive decay chain of \nickel, without evidence of any significant additional energy source, like circumstellar interaction.

The \Siii\ velocity near maximum light is $\sim11,200\kms$, consistent with the normal-velocity class of SNe~Ia as defined by \citet{Wang2009}. The \Siii\ \ld5972 and \ld6355 line pEWs further place \sn\ within the core-normal subclass on the Branch diagram. Overall, the spectroscopic properties of \sn\ are consistent with those of normal SNe~Ia.
Notably, the earliest spectrum at $-10.2$\,days reveals \ion{C}{ii} \ld6580 and \ld7235 features at $\sim13{,}800\kms$. While the \ion{C}{ii} absorption lines are not commonly observed in normal SNe~Ia, it is particularly prevalent among 2003fg-like events.
The slow-evolving light curves paired with luminosity, 
high inferred \nickel\ mass, and the detection of \ion{C}{ii} lines in early spectrum support the classification of \sn\ as a 2003fg-like SN~Ia. Nevertheless, certain characteristics of \sn, such as its position within the normal SNe~Ia parameter space ($t^{i-B}_{max}$ vs. \sbv) on the \cite{Ashall2020} classification diagram, and its relatively large \Siii\ pEWs compared to previously observed 2003fg-like SNe, indicate that \sn\ extends the observed diversity within this peculiar subclass.

On fitting a power-law model, as expected for a homologously expanding ``fireball'', to the early $g$- and $r$-band fluxes yields a time of first light of $t_{\rm first} = -16.43^{+0.45}_{-0.38}$ days relative to $B$-band maximum, with a power-law index of $\alpha = 1.53^{+0.08}_{-0.10}$.
One particularly interesting aspect of \sn\ is that the first photometric detection occurs $1.51^{+0.45}_{-0.38}$ days before (with 99.9\% confidence) the onset of the power-law flux rise. Unlike the flux excesses seen in other SNe~Ia, this behavior cannot be explained solely by interaction with circumstellar material or a companion, which would inevitably produce an additional flux component superimposed on the fireball model's power-law flux rise. Moreover, an emission arising from CSM interaction is expected to show a blue color, whereas in \sn\ the $(g-r)_0$ color of $0.16\pm0.33$\,mag does not show any blue excess, rather consistent with most typical SNe~Ia early colors.
A more plausible explanation for the observed pre-fireball emission is shallow \nickel\ mixing, in which radioactive heating takes place deeper within the ejecta. This leads to a short ``dark phase'', during which the fireball flux rise is delayed by a few days after the explosion. A shallow \nickel\ mixing leading to slower radiation transport can also explain the smaller inferred value of power-law index $\alpha=1.53$, as compared to the classical ``fireball'' model's value of 2. An asymmetric \nickel\ distribution, which is concentrated on the far side of the explosion, may also lead to a similar dark phase.

In principle, a double-detonation model with an early helium flash can produce a short, $1$–$2$ day plateau in the early light curve, as observed in MUSSES1604D \citep{2017Natur.550...80J}. A thin shell of surface helium detonation may account for this early emission and subsequently trigger a detonation in the core of the WD with a few days of delay, thereby explaining the delayed onset of the fireball-like expansion light curve. However, a double-detonation scenario is deemed necessary only for sub-Chandrasekhar mass WDs. In the case of \sn, a 2003fg-like event, with a Chandrasekhar-mass WD already has a sufficiently high central density to initiate carbon ignition. Although it is theoretically possible for a double detonation with surface helium burning to occur in a Chandrasekhar-mass WD, no compelling observational evidence for such a process has been found to date. Instead, this would likely lead to an accretion-induced collapse into a neutron star rather than a thermonuclear explosion. We therefore regard this as a less likely scenario.

Regardless of the physical process behind this pre-fireball flux or early emission, all 2003fg-like SNe with sufficiently early observations -- SNe~2020hvf, 2021zny, 2022ilv, and 2021qvo -- exhibit similar early-time activity. \sn\ thus represents the fifth known event showing such behavior, underscoring the importance of prompt, high-cadence early observations to constrain the physical origin of early emission in these rare SNe~Ia.

Another intriguing aspect of \sn\ is the absence of a clearly associated host galaxy. 2003fg-like SNe~Ia discoveries tend to show a preference towards low-mass dwarf host galaxies \citep[e.g.,][]{2011ApJ...733....3C, 2013MNRAS.432.3117T}, however, these SNe have been found in a range of host-galaxy types, from low-mass dwarfs to massive spiral galaxies \citep{2025A&A...694A..10D, 2025MNRAS.542.2752B}. 
To constrain the host and progenitor of \sn, various possibilities have been thoroughly investigated. The nearest identified galaxy \bose{is an AGN, and} is located at a projected distance of 104\kpc, with a redshift of $z=0.02952$. Considering this as a plausible host, the progenitor could have been a hyper-velocity star (or binary system) ejected from it, traveling far off before the explosion. \bose{An AGN host can facilitate the production of a hyper-velocity progenitor.} The transverse and radial velocity components of such a runaway hyper-velocity star can account for both the large offset from the host galaxy and the redshift difference between the host and that inferred for the SN. Assuming an explosion delay time of $100$\,Myr for the SN, the estimated 3D unprojected velocity of this potential hyper-velocity progenitor is $\sim2200$ km s$^{-1}$.

The second possibility we explore is the presence of an undetected faint dwarf host. Deep GTC imaging obtained 2.5 years after the explosion revealed no host galaxy at the SN location. However, this observation allowed us to place an upper limit on any potential undetected host and to interpret it in the context of the galaxy population. At the $1\sigma$ detection threshold, the point-source limiting magnitude is estimated as $m_{lim,r}=24.4$ mag, and a limiting surface brightness of $\mu_{lim,r}=26.3$ mag arcsec$^{-2}$.
This surface brightness limit rules out the presence of nearly all galaxies within the known luminosity distribution, including most ultra-diffuse galaxies (UDGs), except for those at the faintest end of the UDG luminosity function ($26.3 \lesssim \mu \lesssim 26.5$ mag arcsec$^{-2}$). Nonetheless, the existence of an ultra-faint dwarf galaxy-like (UFDG) host cannot be excluded. Such galaxies, with their small sizes ($\sim30$–70 pc) and extremely low surface brightness ($\mu > 27$ mag arcsec$^{-2}$), could remain undetected even beyond distances of just $\sim20$\,Mpc.

The deep GTC image reveals a faint, diffuse, and extended feature located at a projected distance of approximately $\sim5.5-6.4$\kpc\ from the SN position. Its absolute magnitude is measured as $M_r = -13.2 \pm 1.8$\,mag, with a surface brightness of $\mu_r = 25.58 \pm 0.70$\,mag\,arcsec$^{-2}$. It remains uncertain whether this feature is a galaxy associated with the SN. Nevertheless, its measured surface brightness places it within the luminosity regime of UDGs. 
The feature, however, exhibits an unusually irregular and highly elongated morphology for a UDG in a cluster free environment.
Furthermore, the normalized directional light distance (\ddlr) of \sn\ from this feature is substantially larger than that of the vast majority of observed SNe. A comparison with the ZTF DR2 sample indicates that fewer than $4.0$ percentile of SNe may have potential hosts located at such large separations. Considering these aspects, if this feature is indeed a galaxy, it would be a candidate of significantly low probability, yet can not be ruled out, to be the associated host of \sn. On the other hand, UDGs are low-metallicity systems, which would be a favorable host for a 2003fg-like SNe. If this is the associated host, then \sn\ would be the first SN to be found in any UDG.

Taking into account all these factors, along with the limits we can place on any plausible undetected host galaxy, \sn\ remains one of the strongest candidates for a hostless supernova.
Looking ahead, the combined capabilities of the Vera C. Rubin Observatory’s Legacy Survey of Space and Time (LSST; \citealt{2019ApJ...873..111I}), La Silla Quest Survey 4 (LS4; \citealt{Miller2025}), and The ATLAS Project  \citep{Tonry2018a} will greatly broaden the discovery space for such rare events.

\begin{acknowledgements}

We are grateful to Subo Dong for suggesting our team to followup SN~2021hem, Mark Phillips for discussions, and Morgran Fraser for observational support. 
Funding of the Aarhus supernova group during the creation of this research comes from grants via the Independent Research Fund Denmark (IRFD; grant numbers 8021-00170B and  10.46540/2032-00022B), and  the Aarhus University Research Foundation (Nova project AUFF-E-2023-9-28).
L.G. and C.P.G. recognize the support from the Spanish Ministerio de Ciencia e Innovaci\'on (MCIN) and the Agencia Estatal de Investigaci\'on (AEI) 10.13039/501100011033 under the PID2023-151307NB-I00 SNNEXT project, from Centro Superior de Investigaciones Cient\'ificas (CSIC) under the PIE project 20215AT016 and the program Unidad de Excelencia Mar\'ia de Maeztu CEX2020-001058-M, and from the Departament de Recerca i Universitats de la Generalitat de Catalunya through the 2021-SGR-01270 grant.
C.P.G. acknowledges financial support from the Secretary of Universities and Research (Government of Catalonia) and by the Horizon 2020 Research and Innovation Programme of the European Union under the Marie Sk\l{}odowska-Curie and the Beatriu de Pin\'os 2021 BP 00168 programme.
L.G. acknowledges financial support from AGAUR, CSIC, MCIN and AEI 10.13039/501100011033 under projects PID2023-151307NB-I00, PIE 20215AT016, ILINK23001, COOPB2304.
Based on observations made with the Nordic Optical Telescope, owned in collaboration by the University of Turku and Aarhus University, and operated jointly by Aarhus University, the University of Turku and the University of Oslo, representing Denmark, Finland and Norway, the University of Iceland and Stockholm University at the Observatorio del Roque de los Muchachos, La Palma, Spain, of the Instituto de Astrofisica de Canarias. The NOT data were obtained under programme ID P62-505 and P66-506. 
Based on observations collected at the European Organisation for Astronomical Research in the Southern Hemisphere, Chile, as part of ePESSTO+ (the advanced Public ESO Spectroscopic Survey for Transient Objects Survey – PI: Inserra). ePESSTO+ observations were obtained under ESO program ID 1103.D-0328.
T.E.M.B. is funded by Horizon Europe ERC grant no. 101125877.
J. D. acknowledges support by FCT for CENTRA through the Project No. UIDB/00099/2025. J. D. also acknowledges support by FCT under the PhD grant 2023.01333.BD.
\end{acknowledgements}

\bibliographystyle{aa}
\bibliography{ms.bib}

\begin{appendix}

\onecolumn
\FloatBarrier
\section{Broad-band Photometry of \sn}
\label{sec:optphotometrytable}

Optical and NIR broad-band photometry tables of \sn.

{%
\centering
\fontsize{2.15mm}{3.3mm}\selectfont
\begin{longtable}{c c r c c c c c c c l}
\caption{Optical photometry of \sn.}\label{tab:photsn}\\
\hline\hline
UT Date &        
JD $-$& 
Phase\tablefootmark{a} &  
$B$ &
$V$ &                  
$g$ &                  
$r$ &                  
$i$ &                  
$c$ & 
$o$ & 
Telescope\tablefootmark{b} \\ 
&  
2,459,000    &   
(days) &              
(mag) &              
(mag) &             
(mag) &              
(mag) &             
(mag) &         
(mag)     &       
(mag)     &      
/ Inst.\\
\hline
\endfirsthead
\caption{continued.}\\
\hline\hline
UT Date &        
JD $-$& Phase &  
$B$ &
$V$ &                  
$g$ &                  
$r$ &                  
$i$ &                 
$c$ & 
$o$ & 
Telescope \\ &  
2,459,000    &   
(days) &              
(mag) &              
(mag) &             
(mag) &              
(mag) &             
(mag) &         
(mag)     &      
(mag)     &      
/ Inst. \\
\hline
\endhead
\hline
\endfoot
2021-03-18.47  &   291.97   &   $-$23.6      &      ---      &    ---         & >20.538   &  >20.140   &    ---           &        ---    &      ---         &         ZTF \\
2021-03-22.39  &   295.89   &   $-$19.8      &      ---      &    ---         & >19.600   &  >20.120   &    ---           &        ---    &      ---         &         ZTF \\
2021-03-24.43   &  297.93   &   $-$17.9     &        ---          &        ---          & 20.132 $\pm$ 0.244  & 19.920 $\pm$ 0.211  &        ---          &        ---          &        ---          &             ZTF \\
2021-03-27.50   &  301.00   &   $-$14.9     &        ---          &        ---          & 19.966 $\pm$ 0.263  & 19.656 $\pm$ 0.237  &        ---          &        ---          & 18.156 $\pm$ 0.780  &       ATLAS,ZTF \\
2021-03-28.42   &  301.92   &   $-$14.0     &        ---          &        ---          & 18.968 $\pm$ 0.139  & 18.913 $\pm$ 0.132  &        ---          &        ---          &        ---          &             ZTF \\
2021-03-29.41   &  302.91   &   $-$13.1     &        ---          &        ---          & 18.543 $\pm$ 0.092  & 18.434 $\pm$ 0.076  &        ---          &        ---          &        ---          &             ZTF \\
2021-03-30.46   &  303.96   &   $-$12.1     &        ---          &        ---          & 18.080 $\pm$ 0.058  & 18.036 $\pm$ 0.056  &        ---          &        ---          & 18.173 $\pm$ 0.079  &       ATLAS,ZTF \\
2021-03-31.43   &  304.93   &   $-$11.1     &        ---          &        ---          & 17.724 $\pm$ 0.051  & 17.846 $\pm$ 0.053  &        ---          &        ---          &        ---          &             ZTF \\
2021-04-01.48   &  305.98   &   $-$10.1     &        ---          &        ---          & 17.401 $\pm$ 0.029  &        ---          & 17.680 $\pm$ 0.048  &        ---          & 17.436 $\pm$ 0.024  &       ATLAS,ZTF \\
2021-04-02.45   &  306.95   &    $-$9.2     &        ---          &        ---          & 17.059 $\pm$ 0.026  & 17.060 $\pm$ 0.032  &        ---          &        ---          &        ---          &             ZTF \\
2021-04-03.43   &  307.93   &    $-$8.2     &        ---          &        ---          & 16.809 $\pm$ 0.025  & 16.924 $\pm$ 0.026  &        ---          &        ---          &        ---          &             ZTF \\
2021-04-04.40   &  308.90   &    $-$7.3     &        ---          &        ---          & 16.733 $\pm$ 0.018  & 16.730 $\pm$ 0.029  & 17.137 $\pm$ 0.029  &        ---          &        ---          &             ZTF \\
2021-04-05.47   &  309.97   &    $-$6.3     &        ---          &        ---          & 16.548 $\pm$ 0.021  & 16.664 $\pm$ 0.021  &        ---          &        ---          & 16.776 $\pm$ 0.010  &       ATLAS,ZTF \\
2021-04-06.43   &  310.93   &    $-$5.3     &        ---          &        ---          & 16.427 $\pm$ 0.022  & 16.505 $\pm$ 0.027  &        ---          &        ---          &        ---          &             ZTF \\
2021-04-06.49   &  310.99   &    $-$5.3     &        ---          &        ---          & 16.483 $\pm$ 0.008  &        ---          &        ---          &        ---          &        ---          &             YSE \\
2021-04-07.35   &  311.85   &    $-$4.4     &        ---          &        ---          &        ---          &        ---          & 16.761 $\pm$ 0.029  &        ---          &        ---          &             ZTF \\
2021-04-08.41   &  312.91   &    $-$3.4     &        ---          &        ---          & 16.284 $\pm$ 0.018  & 16.371 $\pm$ 0.024  &        ---          &        ---          &        ---          &             ZTF \\
2021-04-09.32   &  313.82   &    $-$2.5     & 16.362 $\pm$ 0.016  & 16.321 $\pm$ 0.015  &        ---          &        ---          & 16.686 $\pm$ 0.020  &        ---          &        ---          &           LCOGT \\
2021-04-09.55   &  314.05   &    $-$2.3     &        ---          &        ---          &        ---          &        ---          &        ---          & 16.293 $\pm$ 0.006  &        ---          &           ATLAS \\
2021-04-10.06   &  314.56   &    $-$1.8     & 16.486 $\pm$ 0.033  & 16.251 $\pm$ 0.072  &        ---          & 16.392 $\pm$ 0.012  & 16.759 $\pm$ 0.012  &        ---          &        ---          &           LCOGT \\
2021-04-10.44   &  314.94   &    $-$1.5     &        ---          &        ---          &        ---          &        ---          & 16.704 $\pm$ 0.028  &        ---          &        ---          &             ZTF \\
2021-04-11.36   &  315.86   &    $-$0.6     & 16.339 $\pm$ 0.019  & 16.239 $\pm$ 0.018  & 16.139 $\pm$ 0.029  & 16.042 $\pm$ 0.031  & 16.742 $\pm$ 0.028  &        ---          &        ---          &       LCOGT,ZTF \\
2021-04-12.45   &  316.95   &     0.5     & 16.326 $\pm$ 0.017  & 16.255 $\pm$ 0.017  &        ---          &        ---          & 16.766 $\pm$ 0.022  &        ---          &        ---          &           LCOGT \\
2021-04-13.41   &  317.91   &     1.4     &        ---          &        ---          & 16.210 $\pm$ 0.006  & 16.289 $\pm$ 0.006  & 16.719 $\pm$ 0.027  &        ---          &        ---          &         YSE,ZTF \\
2021-04-13.52   &  318.02   &     1.5     &        ---          &        ---          &        ---          &        ---          &        ---          & 16.249 $\pm$ 0.006  &        ---          &           ATLAS \\
2021-04-15.34   &  319.84   &     3.3     & 16.416 $\pm$ 0.020  & 16.274 $\pm$ 0.016  & 16.221 $\pm$ 0.018  & 16.369 $\pm$ 0.015  & 16.732 $\pm$ 0.024  &        ---          &        ---          &       LCOGT,ZTF \\
2021-04-15.49   &  319.99   &     3.4     &        ---          &        ---          &        ---          &        ---          &        ---          &        ---          & 16.425 $\pm$ 0.008  &           ATLAS \\
2021-04-17.44   &  321.94   &     5.3     &        ---          &        ---          &        ---          &        ---          & 16.743 $\pm$ 0.028  & 16.378 $\pm$ 0.007  &        ---          &       ATLAS,ZTF \\
2021-04-17.58   &  322.08   &     5.4     &        ---          &        ---          & 16.331 $\pm$ 0.006  &        ---          & 16.728 $\pm$ 0.008  &        ---          &        ---          &             YSE \\
2021-04-18.33   &  322.83   &     6.1     &        ---          &        ---          & 16.301 $\pm$ 0.023  & 16.343 $\pm$ 0.020  &        ---          &        ---          &        ---          &             ZTF \\
2021-04-19.47   &  323.97   &     7.2     &        ---          &        ---          &        ---          &        ---          &        ---          &        ---          & 16.553 $\pm$ 0.009  &           ATLAS \\
2021-04-20.35   &  324.85   &     8.1     &        ---          &        ---          & 16.447 $\pm$ 0.021  & 16.404 $\pm$ 0.025  & 16.935 $\pm$ 0.028  &        ---          &        ---          &             ZTF \\
2021-04-22.46   &  326.96   &    10.1     &        ---          &        ---          &        ---          & 16.575 $\pm$ 0.007  &        ---          &        ---          &        ---          &             YSE \\
2021-04-23.11   &  327.61   &    10.8     & 16.887 $\pm$ 0.022  & 16.537 $\pm$ 0.016  &        ---          & 16.756 $\pm$ 0.021  & 17.017 $\pm$ 0.027  &        ---          &        ---          &           LCOGT \\
2021-04-23.47   &  327.97   &    11.1     &        ---          &        ---          &        ---          &        ---          &        ---          &        ---          & 16.750 $\pm$ 0.014  &           ATLAS \\
2021-04-24.38   &  328.88   &    12.0     &        ---          &        ---          & 16.737 $\pm$ 0.023  & 16.636 $\pm$ 0.026  &        ---          &        ---          & 16.821 $\pm$ 0.020  &       ATLAS,ZTF \\
2021-04-25.01   &  329.51   &    12.6     & 16.568 $\pm$ 0.653  &        ---          &        ---          &        ---          &        ---          &        ---          &        ---          &           LCOGT \\
2021-04-25.41   &  329.91   &    13.0     &        ---          &        ---          & 16.743 $\pm$ 0.024  &        ---          & 16.999 $\pm$ 0.035  &        ---          & 16.848 $\pm$ 0.019  &       ATLAS,ZTF \\
2021-04-27.48   &  331.98   &    15.0     &        ---          &        ---          &        ---          &        ---          &        ---          &        ---          & 16.885 $\pm$ 0.061  &           ATLAS \\
2021-04-29.38   &  333.88   &    16.8     & 17.573 $\pm$ 0.062  & 16.949 $\pm$ 0.034  &        ---          &        ---          & 17.033 $\pm$ 0.038  &        ---          & 16.906 $\pm$ 0.015  &     ATLAS,LCOGT \\
2021-04-30.42   &  334.92   &    17.8     &        ---          &        ---          & 17.245 $\pm$ 0.025  & 16.790 $\pm$ 0.026  & 17.074 $\pm$ 0.031  &        ---          &        ---          &             ZTF \\
2021-05-01.41   &  335.91   &    18.8     &        ---          &        ---          & 17.370 $\pm$ 0.031  & 16.869 $\pm$ 0.026  &        ---          &        ---          & 16.977 $\pm$ 0.031  &       ATLAS,ZTF \\
2021-05-02.40   &  336.90   &    19.7     &        ---          &        ---          & 17.404 $\pm$ 0.027  & 16.826 $\pm$ 0.026  &        ---          &        ---          &        ---          &             ZTF \\
2021-05-02.61   &  337.11   &    19.9     &        ---          &        ---          &        ---          & 16.907 $\pm$ 0.009  & 17.052 $\pm$ 0.010  &        ---          &        ---          &             YSE \\
2021-05-03.19   &  337.69   &    20.5     & 18.059 $\pm$ 0.029  & 17.175 $\pm$ 0.019  &        ---          &        ---          & 17.034 $\pm$ 0.024  &        ---          &        ---          &           LCOGT \\
2021-05-03.52   &  338.02   &    20.8     &        ---          &        ---          &        ---          &        ---          &        ---          &        ---          & 16.985 $\pm$ 0.012  &           ATLAS \\
2021-05-05.38   &  339.88   &    22.6     &        ---          &        ---          & 17.722 $\pm$ 0.030  & 16.896 $\pm$ 0.029  &        ---          & 17.463 $\pm$ 0.065  &        ---          &       ATLAS,ZTF \\
2021-05-06.40   &  340.90   &    23.6     &        ---          &        ---          & 17.886 $\pm$ 0.031  & 17.002 $\pm$ 0.024  &        ---          &        ---          &        ---          &             ZTF \\
2021-05-07.36   &  341.86   &    24.5     &        ---          &        ---          & 17.970 $\pm$ 0.033  & 17.124 $\pm$ 0.022  &        ---          &        ---          &        ---          &             ZTF \\
2021-05-07.58   &  342.08   &    24.7     &        ---          &        ---          &        ---          &        ---          &        ---          &        ---          & 17.107 $\pm$ 0.016  &           ATLAS \\
2021-05-08.37   &  342.87   &    25.5     &        ---          &        ---          & 18.066 $\pm$ 0.034  &        ---          &        ---          &        ---          &        ---          &             ZTF \\
2021-05-08.55   &  343.05   &    25.7     &        ---          &        ---          & 18.051 $\pm$ 0.015  & 17.198 $\pm$ 0.008  &        ---          &        ---          &        ---          &             YSE \\
2021-05-09.46   &  343.96   &    26.5     &        ---          &        ---          & 18.122 $\pm$ 0.036  &        ---          &        ---          & 17.693 $\pm$ 0.022  &        ---          &       ATLAS,ZTF \\
2021-05-10.36   &  344.86   &    27.4     &        ---          &        ---          &        ---          &        ---          & 17.151 $\pm$ 0.035  &        ---          &        ---          &             ZTF \\
2021-05-11.29   &  345.79   &    28.3     & 18.845 $\pm$ 0.106  & 17.707 $\pm$ 0.041  & 18.280 $\pm$ 0.037  & 17.232 $\pm$ 0.031  & 17.316 $\pm$ 0.076  &        ---          &        ---          &       LCOGT,ZTF \\
2021-05-11.49   &  345.99   &    28.5     &        ---          &        ---          & 18.297 $\pm$ 0.016  &        ---          & 17.280 $\pm$ 0.008  &        ---          & 17.318 $\pm$ 0.015  &       ATLAS,YSE \\
2021-05-12.38   &  346.88   &    29.4     & 18.936 $\pm$ 0.015  & 17.850 $\pm$ 0.012  & 18.367 $\pm$ 0.039  & 17.427 $\pm$ 0.010  &        ---          &        ---          &        ---          &       LCOGT,ZTF \\
2021-05-13.47   &  347.97   &    30.4     &        ---          &        ---          &        ---          &        ---          &        ---          & 17.876 $\pm$ 0.022  &        ---          &           ATLAS \\
2021-05-14.29   &  348.79   &    31.2     & 18.922 $\pm$ 0.077  & 17.851 $\pm$ 0.064  &        ---          &        ---          & 17.469 $\pm$ 0.027  &        ---          &        ---          &           LCOGT \\
2021-05-15.37   &  349.87   &    32.2     &        ---          &        ---          & 18.513 $\pm$ 0.043  & 17.455 $\pm$ 0.031  & 17.422 $\pm$ 0.040  &        ---          & 17.514 $\pm$ 0.018  &       ATLAS,ZTF \\
2021-05-17.41   &  351.91   &    34.2     &        ---          &        ---          &        ---          &        ---          &        ---          & 18.088 $\pm$ 0.025  &        ---          &           ATLAS \\
2021-05-18.34   &  352.84   &    35.1     &        ---          &        ---          & 18.597 $\pm$ 0.046  & 17.678 $\pm$ 0.033  & 17.530 $\pm$ 0.041  &        ---          &        ---          &             ZTF \\
2021-05-19.32   &  353.82   &    36.1     & 19.118 $\pm$ 0.042  & 18.121 $\pm$ 0.023  &        ---          & 17.649 $\pm$ 0.032  & 17.767 $\pm$ 0.028  &        ---          & 17.727 $\pm$ 0.014  & ATLAS,LCOGT,ZTF \\
2021-05-19.51   &  354.01   &    36.2     &        ---          &        ---          & 18.606 $\pm$ 0.027  &        ---          & 17.687 $\pm$ 0.014  &        ---          &        ---          &             YSE \\
2021-05-20.38   &  354.88   &    37.1     &        ---          &        ---          & 18.695 $\pm$ 0.053  & 17.750 $\pm$ 0.032  &        ---          &        ---          &        ---          &             ZTF \\
2021-05-21.43   &  355.93   &    38.1     &        ---          &        ---          &        ---          & 17.846 $\pm$ 0.039  & 17.827 $\pm$ 0.072  & 18.145 $\pm$ 0.030  &        ---          &       ATLAS,ZTF \\
2021-05-22.34   &  356.84   &    39.0     &        ---          &        ---          & 18.596 $\pm$ 0.057  & 17.763 $\pm$ 0.036  &        ---          &        ---          &        ---          &             ZTF \\
2021-05-23.40   &  357.90   &    40.0     &        ---          &        ---          &        ---          &        ---          &        ---          &        ---          & 17.883 $\pm$ 0.026  &           ATLAS \\
2021-05-24.37   &  358.87   &    40.9     &        ---          &        ---          & 18.667 $\pm$ 0.091  & 17.834 $\pm$ 0.046  &        ---          &        ---          & 17.835 $\pm$ 0.048  &       ATLAS,ZTF \\
2021-05-25.39   &  359.89   &    41.9     &        ---          &        ---          & 18.668 $\pm$ 0.107  &        ---          & 17.854 $\pm$ 0.064  &        ---          &        ---          &             ZTF \\
2021-05-26.03   &  360.53   &    42.5     & 19.331 $\pm$ 1.136  &        ---          &        ---          &        ---          & 17.952 $\pm$ 0.488  &        ---          &        ---          &           LCOGT \\
2021-05-27.31   &  361.81   &    43.8     &        ---          &        ---          & 18.803 $\pm$ 0.098  & 17.895 $\pm$ 0.046  &        ---          &        ---          &        ---          &             ZTF \\
2021-05-27.45   &  361.95   &    43.9     &        ---          &        ---          & 18.788 $\pm$ 0.117  &        ---          &        ---          &        ---          &        ---          &             ZTF \\
2021-05-27.54   &  362.04   &    44.0     &        ---          &        ---          &        ---          &        ---          &        ---          &        ---          & 17.917 $\pm$ 0.168  &           ATLAS \\
2021-05-28.31   &  362.81   &    44.7     &        ---          &        ---          & 18.679 $\pm$ 0.076  & 17.893 $\pm$ 0.042  & 17.965 $\pm$ 0.056  &        ---          &        ---          &             ZTF \\
2021-05-29.35   &  363.85   &    45.7     &        ---          &        ---          & 18.755 $\pm$ 0.072  & 17.966 $\pm$ 0.040  &        ---          &        ---          &        ---          &             ZTF \\
2021-05-29.40   &  363.90   &    45.8     &        ---          &        ---          &        ---          &        ---          &        ---          &        ---          & 18.137 $\pm$ 0.045  &           ATLAS \\
2021-05-29.50   &  364.00   &    45.9     &        ---          &        ---          &        ---          & 18.030 $\pm$ 0.023  & 18.094 $\pm$ 0.023  &        ---          &        ---          &             YSE \\
2021-05-30.37   &  364.87   &    46.7     &        ---          &        ---          &        ---          & 17.959 $\pm$ 0.041  &        ---          &        ---          & 18.115 $\pm$ 0.031  &       ATLAS,ZTF \\
2021-05-30.98   &  365.48   &    47.3     & 19.200 $\pm$ 0.086  & 18.379 $\pm$ 0.041  &        ---          &        ---          & 18.151 $\pm$ 0.038  &        ---          &        ---          &           LCOGT \\
2021-05-31.35   &  365.85   &    47.7     &        ---          &        ---          & 18.853 $\pm$ 0.059  & 18.020 $\pm$ 0.039  & 18.110 $\pm$ 0.056  &        ---          &        ---          &             ZTF \\
2021-06-01.34   &  366.84   &    48.6     &        ---          &        ---          & 18.922 $\pm$ 0.059  & 18.060 $\pm$ 0.041  &        ---          &        ---          &        ---          &             ZTF \\
2021-06-02.36   &  367.86   &    49.6     &        ---          &        ---          & 18.874 $\pm$ 0.062  & 18.120 $\pm$ 0.042  &        ---          &        ---          & 18.209 $\pm$ 0.033  &       ATLAS,ZTF \\
2021-06-03.31   &  368.81   &    50.5     &        ---          &        ---          & 18.924 $\pm$ 0.058  & 18.161 $\pm$ 0.041  & 18.194 $\pm$ 0.059  &        ---          &        ---          &             ZTF \\
2021-06-04.31   &  369.81   &    51.5     & 19.471 $\pm$ 0.037  & 18.495 $\pm$ 0.025  & 18.895 $\pm$ 0.057  & 18.162 $\pm$ 0.042  & 18.391 $\pm$ 0.029  & 18.463 $\pm$ 0.026  &        ---          & ATLAS,LCOGT,ZTF \\
2021-06-06.34   &  371.84   &    53.4     &        ---          &        ---          & 18.987 $\pm$ 0.058  & 18.207 $\pm$ 0.042  &        ---          &        ---          &        ---          &             ZTF \\
2021-06-06.41   &  371.91   &    53.5     &        ---          &        ---          &        ---          &        ---          &        ---          &        ---          & 18.275 $\pm$ 0.033  &           ATLAS \\
2021-06-06.50   &  372.00   &    53.6     &        ---          &        ---          & 18.942 $\pm$ 0.047  &        ---          & 18.340 $\pm$ 0.022  &        ---          &        ---          &             YSE \\
2021-06-08.19   &  373.69   &    55.2     &        ---          &        ---          &        ---          &        ---          & 18.491 $\pm$ 0.076  &        ---          &        ---          &             ZTF \\
2021-06-09.47   &  374.97   &    56.5     &        ---          &        ---          & 19.102 $\pm$ 0.047  & 18.321 $\pm$ 0.025  &        ---          &        ---          &        ---          &             YSE \\
2021-06-10.33   &  375.83   &    57.3     &        ---          &        ---          & 18.950 $\pm$ 0.060  & 18.398 $\pm$ 0.049  &        ---          &        ---          &        ---          &             ZTF \\
2021-06-11.30   &  376.80   &    58.2     &        ---          &        ---          & 18.991 $\pm$ 0.062  & 18.364 $\pm$ 0.047  &        ---          &        ---          &        ---          &             ZTF \\
2021-06-12.38   &  377.88   &    59.3     &        ---          &        ---          & 19.042 $\pm$ 0.062  & 18.417 $\pm$ 0.049  &        ---          & 18.693 $\pm$ 0.033  &        ---          &       ATLAS,ZTF \\
2021-06-13.19   &  378.69   &    60.1     & 19.543 $\pm$ 0.033  & 18.687 $\pm$ 0.025  &        ---          &        ---          & 18.696 $\pm$ 0.026  &        ---          &        ---          &           LCOGT \\
2021-06-13.28   &  378.78   &    60.1     &        ---          &        ---          & 19.044 $\pm$ 0.062  & 18.437 $\pm$ 0.050  & 18.582 $\pm$ 0.074  &        ---          &        ---          &             ZTF \\
2021-06-13.45   &  378.95   &    60.3     &        ---          &        ---          & 19.046 $\pm$ 0.030  &        ---          & 18.584 $\pm$ 0.021  &        ---          &        ---          &             YSE \\
2021-06-14.33   &  379.83   &    61.2     &        ---          &        ---          & 19.051 $\pm$ 0.071  & 18.513 $\pm$ 0.052  &        ---          &        ---          &        ---          &             ZTF \\
2021-06-15.08   &  380.58   &    61.9     & 19.569 $\pm$ 0.052  & 18.370 $\pm$ 0.208  &        ---          &        ---          &        ---          &        ---          &        ---          &           LCOGT \\
2021-06-16.33   &  381.83   &    63.1     &        ---          &        ---          & 18.976 $\pm$ 0.105  & 18.580 $\pm$ 0.116  & 18.718 $\pm$ 0.113  &        ---          &        ---          &             ZTF \\
2021-06-16.45   &  381.95   &    63.2     &        ---          &        ---          & 19.105 $\pm$ 0.040  & 18.551 $\pm$ 0.031  &        ---          &        ---          &        ---          &             YSE \\
2021-06-17.29   &  382.79   &    64.0     &        ---          &        ---          & 19.075 $\pm$ 0.087  & 18.511 $\pm$ 0.060  &        ---          &        ---          &        ---          &             ZTF \\
2021-06-18.33   &  383.83   &    65.0     &        ---          &        ---          & 19.082 $\pm$ 0.097  & 18.543 $\pm$ 0.072  &        ---          &        ---          & 18.525 $\pm$ 0.048  &       ATLAS,ZTF \\
2021-06-18.81   &  384.31   &    65.5     & 19.614 $\pm$ 0.106  & 18.905 $\pm$ 0.054  &        ---          &        ---          & 18.873 $\pm$ 0.063  &        ---          &        ---          &           LCOGT \\
2021-06-19.43   &  384.93   &    66.1     &        ---          &        ---          & 19.211 $\pm$ 0.053  &        ---          & 18.772 $\pm$ 0.026  &        ---          & 18.868 $\pm$ 0.179  &       ATLAS,YSE \\
2021-06-20.37   &  385.87   &    67.0     &        ---          &        ---          & 19.138 $\pm$ 0.105  & 18.660 $\pm$ 0.071  &        ---          &        ---          & 18.739 $\pm$ 0.083  &       ATLAS,ZTF \\
2021-06-21.37   &  386.87   &    67.9     &        ---          &        ---          &        ---          &        ---          &        ---          &        ---          & 18.661 $\pm$ 0.109  &           ATLAS \\
2021-06-22.19   &  387.69   &    68.7     & 19.785 $\pm$ 0.236  & 18.804 $\pm$ 0.096  & 19.247 $\pm$ 0.163  & 18.593 $\pm$ 0.087  & 18.912 $\pm$ 0.132  &        ---          &        ---          &       LCOGT,ZTF \\
2021-06-22.43   &  387.93   &    69.0     &        ---          &        ---          &        ---          &        ---          &        ---          &        ---          & 18.967 $\pm$ 0.155  &           ATLAS \\
2021-06-23.38   &  388.88   &    69.9     &        ---          &        ---          &        ---          &        ---          &        ---          &        ---          & 18.900 $\pm$ 0.132  &           ATLAS \\
2021-06-24.36   &  389.86   &    70.8     &        ---          &        ---          &        ---          &        ---          &        ---          &        ---          & 18.942 $\pm$ 0.127  &           ATLAS \\
2021-06-24.87   &  390.37   &    71.3     & 19.573 $\pm$ 0.274  & 18.932 $\pm$ 0.136  &        ---          &        ---          & 18.785 $\pm$ 0.142  &        ---          &        ---          &           LCOGT \\
2021-06-25.40   &  390.90   &    71.8     &        ---          &        ---          &        ---          &        ---          &        ---          &        ---          & 18.409 $\pm$ 0.143  &           ATLAS \\
2021-06-26.31   &  391.81   &    72.7     &        ---          &        ---          &        ---          & 18.803 $\pm$ 0.081  & 19.005 $\pm$ 0.134  &        ---          &        ---          &             ZTF \\
2021-06-26.38   &  391.88   &    72.8     &        ---          &        ---          &        ---          &        ---          &        ---          &        ---          & 18.886 $\pm$ 0.077  &           ATLAS \\
2021-06-26.41   &  391.91   &    72.8     &        ---          &        ---          &        ---          & 18.811 $\pm$ 0.034  & 18.967 $\pm$ 0.032  &        ---          &        ---          &             YSE \\
2021-06-27.85   &  393.35   &    74.2     & 19.777 $\pm$ 0.127  & 19.109 $\pm$ 0.076  &        ---          &        ---          & 19.128 $\pm$ 0.087  &        ---          &        ---          &           LCOGT \\
2021-06-28.35   &  393.85   &    74.7     &        ---          &        ---          & 19.139 $\pm$ 0.080  & 18.927 $\pm$ 0.073  &        ---          &        ---          & 19.027 $\pm$ 0.061  &       ATLAS,ZTF \\
2021-06-29.25   &  394.75   &    75.5     &        ---          &        ---          &        ---          &        ---          & 18.964 $\pm$ 0.116  &        ---          &        ---          &             ZTF \\
2021-06-30.27   &  395.77   &    76.5     &        ---          &        ---          & 19.218 $\pm$ 0.159  & 18.739 $\pm$ 0.189  &        ---          &        ---          &        ---          &             ZTF \\
2021-06-30.41   &  395.91   &    76.7     &        ---          &        ---          &        ---          & 18.936 $\pm$ 0.031  & 19.139 $\pm$ 0.037  &        ---          &        ---          &             YSE \\
2021-07-02.35   &  397.85   &    78.5     &        ---          &        ---          & 19.406 $\pm$ 0.087  &        ---          & 19.131 $\pm$ 0.134  &        ---          & 19.106 $\pm$ 0.054  &       ATLAS,ZTF \\
2021-07-06.37   &  401.87   &    82.4     &        ---          &        ---          &        ---          &        ---          & 19.309 $\pm$ 0.034  &        ---          & 19.327 $\pm$ 0.065  &       ATLAS,YSE \\
2021-07-08.38   &  403.88   &    84.4     &        ---          &        ---          &        ---          &        ---          &        ---          & 19.227 $\pm$ 0.059  &        ---          &           ATLAS \\
2021-07-10.34   &  405.84   &    86.2     &        ---          &        ---          &        ---          &        ---          &        ---          &        ---          & 19.237 $\pm$ 0.068  &           ATLAS \\
2021-07-13.40   &  408.90   &    89.2     &        ---          &        ---          & 19.516 $\pm$ 0.045  &        ---          & 19.466 $\pm$ 0.042  &        ---          &        ---          &             YSE \\
2021-07-14.33   &  409.83   &    90.1     &        ---          &        ---          &        ---          &        ---          &        ---          &        ---          & 19.450 $\pm$ 0.070  &           ATLAS \\
\end{longtable}

\begin{longtable}{c c r c c l}
\caption{NIR photometry of \sn.}\label{tab:photsnNIR}\\
\hline\hline
UT Date &        
JD $-$& 
Phase\tablefootmark{a} &  
$J$ &
$H$ &  
Telescope\tablefootmark{b} \\ 
&  
2,459,000    &   
(days) &              
(mag) &              
(mag) &      
/ Inst.\\
\hline
\endfirsthead
\caption{continued.}\\
\hline\hline
UT Date &        
JD $-$& Phase &  
$J$ &
$H$ &       
Telescope \\ &  
2,459,000    &   
(days) &              
(mag) &              
(mag) &       
/ Inst. \\
\hline
\endhead
\hline
\endfoot

2021-04-10.42  &   2459314.92   &    -1.5   &   16.409 $\pm$ 0.033 &  16.602 $\pm$ 0.044  &  NOT \\
2021-04-21.18  &   2459325.68   &     8.9   &   16.790 $\pm$ 0.027 &  16.628 $\pm$ 0.020  &  NOT \\
2021-05-11.22  &   2459345.72   &    28.2   &   16.953 $\pm$ 0.041 &  16.639 $\pm$ 0.032  &  NOT\\
\end{longtable}

\tablefoot{\tablefoottext{a}{Rest-frame days relative to the epoch of $B$-band maximum, i.e., \PeakEpoch.}
\tablefoottext{b}{The abbreviations for the telescope/instrument are: 
LCOGT - Las Cumbres Observatory telescope network observatons through FLOWS survey; 
NOT - Nordic Optical Telescope; 
ATLAS - ATLAS survey telescope's orange and cyan filter observations; 
YSE - Pan-STARRS survey telescope observation through YSE survey; ZTF - ZTF survey telescopes. Data observed within 5\,hr are represented under a single-epoch observation.}
}

}

\section{}
\FloatBarrier

\begin{figure}[!ht]
\centering
	\includegraphics[width=0.45\linewidth]{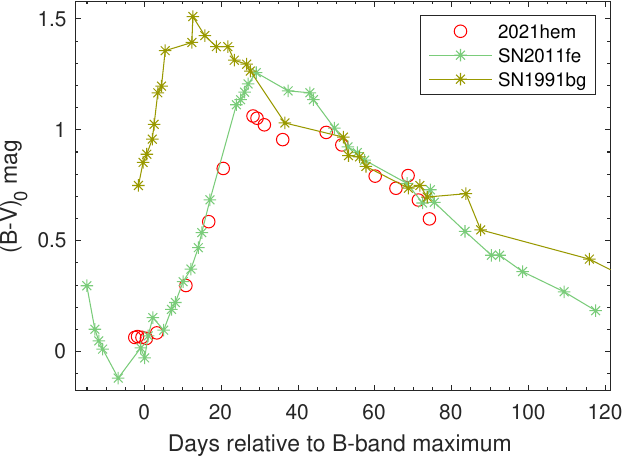}
	\caption{Intrinsic $(\bv)_0$ color curve of \sn\ corrected only for Milky-Way reddening is shown, and is compared with normal SN~Ia 2011fe, and low-luminosity SN~Ia 1991bg.}
	\label{fig:color_comp}
\end{figure}

\begin{figure}[!ht]
\centering
	\includegraphics[width=0.4\linewidth]{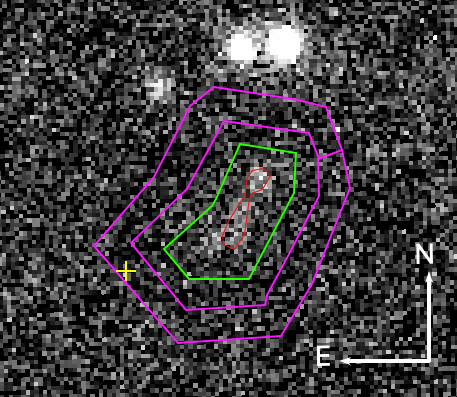}
	\caption{Photometry region of the diffused extended feature at 4.2\kpc\ distance from SN location (yellow cross). The green polygon region is the aperture used for photometry, and the magenta polygon annulus represents the region for background estimation. The red contour represents the bright region of the feature used for estimating the surface brightness.}
	\label{fig:gal_phot_reg}
\end{figure}

\onecolumn
\section{Probabilistic Multi-dimensional Clustering Analysis} \label{sec:branch}

As previously discussed in Sect.~\ref{sec:results}, \sn\ possesses many 
of the characteristics of 2003fg-like SNe discussed in \citealt{Ashall2021}, 
(see their Fig. 18). As shown there, most of the 2003fg-like objects are 
classified as ``shallow-silicon'' objects within the Branch scheme 
\citep{Branch2006}, with a couple objects classified as ``core-normals''. 
\citet{Burrow2020} re-formulate the Branch Diagram 
using Gaussian Mixture Models, providing a method by which to estimate 
the probability any individual SN~Ia belongs to each Branch group. Their
clustering analysis also extended beyond the Branch Diagram phase
space (i.e., pEW \ion{Si}{ii}\ $\lambda5972$ vs. pEW \ion{Si}{ii}\ $\lambda6355)$ to 
include $v_{\ion{Si}{ii}}$ and $M_B$ in both three- and four-dimensional
spaces. While the Branch groups are robust and tightly constrained
even in higher-dimensional spaces, the groupings in these higher-dimensional 
spaces are by definition distinct from the original Branch groups.

\begin{figure}[h!]
    \centering
    \includegraphics[height=3in]{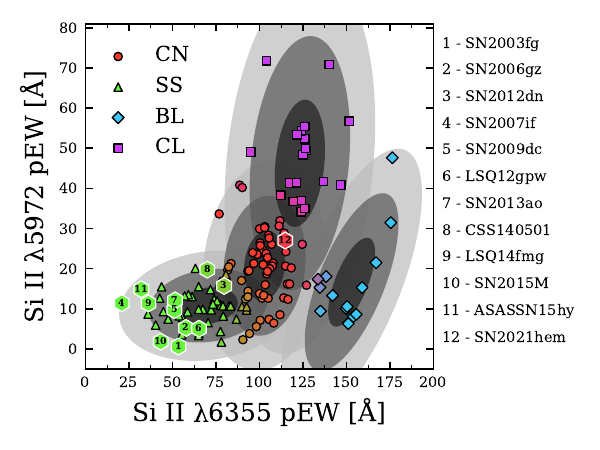}
    \caption{Branch Diagram including \sn\ (assuming $z = 0.0363$), 
    the sample of 2003fg-like  SNe~Ia from \citet{Ashall2021} over-plotted 
    with the \citet{Burrow2020} sample. We note that both CSS140501 and 
    SN~2012dn have both moved from the ``core-normal'' classifications 
    assigned in \citet{Ashall2021} to the ``shallow-silicon'' groups 
    here. See Table~\ref{tab:03fg_branch2D} for the full probabilities,
    which correspond to the colorings of the overplotted points.}
    \label{fig:branch2D}
\end{figure}

Using the measurements from \citet{Ashall2021} and those of \sn\
determined in Sect.~\ref{sec:specanal}, we re-plot the traditional (e.g. 
two-dimensional) Branch Diagram in Fig.~\ref{fig:branch2D}. Because 
group membership is determined probabilistically, we find that all 
2003fg-likes now belong to the ``shallow-silicon'' group, with a 
minimum membership probability of $p > 0.79$ (see Table~\ref{tab:03fg_branch2D} 
for full probabilities). We find \sn\ is classified as a ``core-normal''
with $p = 0.8526$, significantly offset from the rest of the 2003fg-like 
sample. This is consistent with the placement of \sn\ in the \sbv\ vs.
$t^{i-B}_{max}$ space defined in \citet{Ashall2021} where it sits 
outside the region of 2003fg-like SNe~Ia.

\begin{table}[h!]
    \centering
    \caption{Branch Group Membership Probabilities of 2003fg-like SNe~Ia}
    \label{tab:03fg_branch2D}
    \begin{tabular}{c|cccc}
        \hline\hline
        SN Name& CN & SS & BL & CL \\
        \hline
        SN~2003fg & \scinot{4.147}{-4} & 0.9996 & \scinot{1.207}{-16} & \scinot{3.758}{-6} \\
        SN~2006gz & \scinot{6.572}{-4} & 0.9993 & \scinot{5.780}{-17} & \scinot{3.908}{-6} \\
        SN~2012dn & 0.2025 & 0.7962 & \scinot{2.515}{-12} & 0.0013 \\
        SN~2007if & \scinot{2.490}{-9} & 0.9999 & \scinot{2.834}{-34} & \scinot{1.453}{-10} \\
        SN~2009dc & \scinot{9.233}{-5} & 0.9999 & \scinot{6.830}{-21} & \scinot{5.998}{-7} \\
        LSQ12gpw & 0.0048 & 0.9952 & \scinot{2.368}{-14} & \scinot{2.594}{-5} \\
        SN~2013ao & \scinot{1.376}{-4} & 0.9999 & \scinot{1.906}{-21} & \scinot{8.968}{-7} \\
        CSS140501 & 0.1244 & 0.8748 & \scinot{3.976}{-16} & \scinot{8.772}{-4} \\
        LSQ14fmg & \scinot{8.573}{-7} & 0.9999 & \scinot{2.002}{-27} & \scinot{1.273}{-8} \\
        SN~2015M & \scinot{1.272}{-5} & 0.9999 & \scinot{6.699}{-21} & \scinot{1.606}{-7} \\
        ASASSN15hy & \scinot{4.136}{-7} & 0.9999 & \scinot{2.955}{-30} & \scinot{9.391}{-9} \\
        SN~2021hem & 0.8526 & \scinot{3.751}{-4} & \scinot{5.936}{-6} & 0.1470 \\
        \hline
    \end{tabular}
\end{table}

Extending the above analysis to higher dimensional spaces by 
including $v_{\ion{Si}{ii}}$ and $M_B$ in the GMM clustering 
requires addressing the uncertainties in both quantities 
associated with the seeming hostless nature of \sn. We examine 
3 cases: 
(1) assuming \sn\ lies at $z=0.0363$ as adopted in the main 
body of the paper; 
(2) the runaway progenitor case which assuming \sn\ lies within 
2MASX~J16212572{+}1431537 at $z=0.02952$; 
and (3) assuming $z=0.0363$ but with larger uncertainties derived
from propagating the redshift uncertainty as discussed in Sect.~\ref{sec:redshift}. 
We note that cases (1) and (2) consider only the measurement uncertainties
associated with the \ion{Si}{ii} pEWs, $v_{\rm Si~II}$, and $M_B$. The results 
of this analysis are shown below in Fig. \ref{fig:multid} 
and Table \ref{tab:prob_multid}.

We find that that despite the overluminous nature of \sn, the 
inclusion of $M_B$ in both cases (1) and (3) changes the membership
of \sn\ from the  ``core-normal''-equivalent group to the 
``cool''-equivalent group. The inclusion of $v_{\rm Si~II}$ causes the
probabilities to shift 20\% in the same direction without changing the 
group membership of \sn. Given the well-established association
between underluminous, rapidly-declining SNe~Ia and the ``Cool'' Branch 
group \citep{Branch2006,Burrow2020,Phillips2025}; the placement of the 
overluminous, slowly-declining \sn\ in the ``cool'' group is likely a reflection 
of the abnormal, hostless nature of \sn, but may hint at our incomplete 
understanding of of 2003fg-like SNe~Ia. This placement likely arises due to:
(1) a tight clustering of the ``core-normal'' objects along the $v_{\rm Si~II}$ 
and $M_B$ axes,
(2) an upper-limit on pseudo-equivalent widths of the Si~II lines for 
membership in the ``shallow-silicon'' group,
and (3) the ``broad-line'' group being defined by high photospheric velocities
to the point of being almost completely distinct from the other Branch groups in
the higher dimensional spaces \citep{Burrow2020}.

In case (2), we find that \sn\ is associated with the ``core-normal''-equivalent
group at $p > 0.8$ in all dimensionalities. \sn 's placement outside of the 
``shallow-silicon'' (or equivalent) group in all cases and dimensionalities 
suggests that within this clustering scheme, its 
properties are significantly different than expected for a 2003fg-like SNe~Ia
at $z=0.0363$. Instead, this method of analysis, taken in combination with 
the location of \sn\ in the \citet{Ashall2020} classification, is more 
consistent with \sn\ being a overluminous ``core-normal'' SN~Ia arising 
from the hyper-velocity progenitor scenario discussed in Sect.~\ref{sec:arunawayburnoutnooneloved}. However, 
given both the diversity of 2003fg-like SNe~Ia behavior and the hostless nature of
\sn, such results should be evaluated in combination with other analyses and
remembering the large uncertainties on all measurements.

\begin{figure}[h!]
    \centering
    \includegraphics[width=\columnwidth]{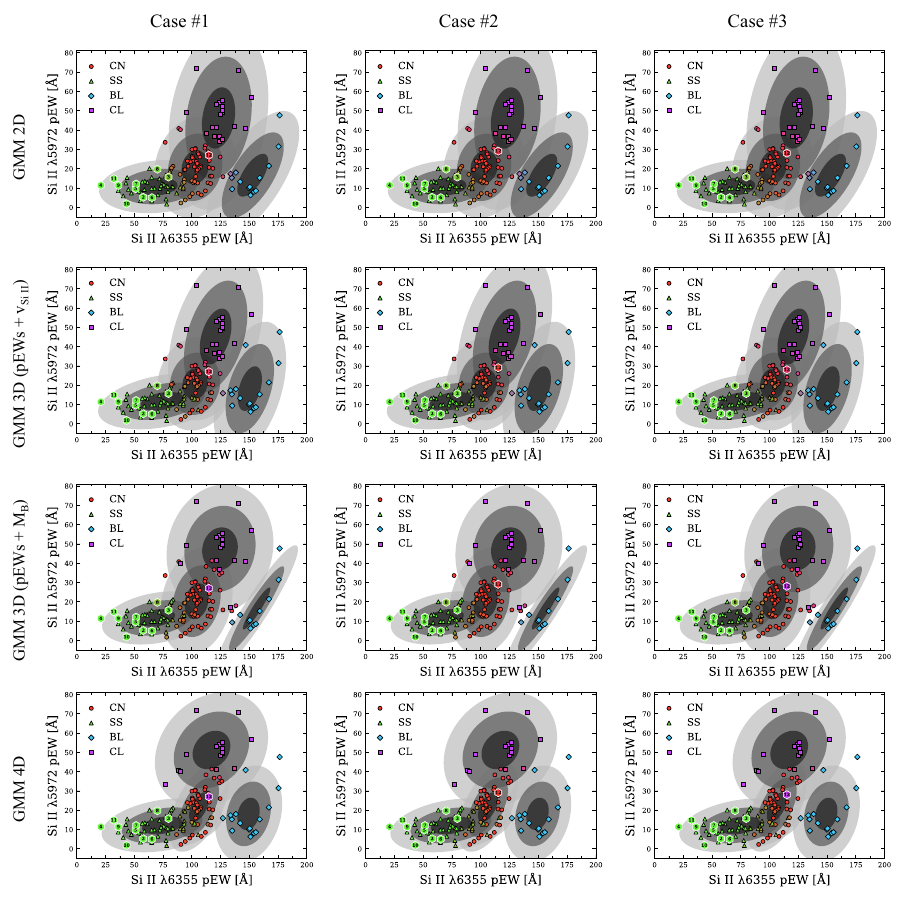}
    \caption{Multi-dimensional clusterings projected into the Branch
    Diagram phase-space for all combinations of input data and cluster
    dimensionality. The full probabilities associated with the group
    membership of \sn\ can be found in Table \ref{tab:prob_multid}.
    Numerical labels on the overplotted data are the same as in Fig. 
    \ref{fig:branch2D}.}
    \label{fig:multid}
\end{figure}

\begin{table}[h!]
    \centering
    \caption{Multi-dimensional ``Branch'' Group Membership Probabilities of \sn}
    \label{tab:prob_multid}
    \begin{tabular}{c|cccc}
        \hline\hline
         & CN & SS & BL & CL \\
        \hline
        \noalign{\smallskip}
        \multicolumn{5}{c}{Case 1: $z=0.0363$} \\
        \noalign{\smallskip}
        \hline
        2D (Branch Diagram) & 0.8526 & 0.0004 & \scinot{5.936}{-5} & 0.1470 \\
        3D (pEWs+$v_{Si~II}$) & 0.6847 & 0.0081 & 0.0017 & 0.3054 \\
        3D (pEWs+$M_B$) & 0.0098 & 0.0018 & \scinot{4.621}{-54} & 0.9884 \\
        4D (pEWs+$v_{Si~II}$+$M_B$) & 0.0057 & 0.0066 & \scinot{6.914}{-10} & 0.9876 \\
        \hline
        \noalign{\smallskip}
        \multicolumn{5}{c}{Case 2: Runaway Progenitor ($z=0.02952$)} \\
        \noalign{\smallskip}
        \hline
        2D (Branch Diagram) & 0.8018 & \scinot{8.450}{-5} & \scinot{3.123}{-6} & 0.1981 \\
        3D (pEWs+$v_{Si~II}$) & 0.9893 & 0.0037 & \scinot{1.508}{-5} & 0.0070 \\
        3D (pEWs+$M_B$) & 0.9011 & 0.0006 & \scinot{4.304}{-44} & 0.0983 \\
        4D (pEWs+$v_{Si~II}$+$M_B$) & 0.9959 & 0.0022 & \scinot{4.902}{-15} & 0.0019 \\
        \hline
        \noalign{\smallskip}
        \multicolumn{5}{c}{Case 3: $z=0.0363$ with propagated uncertainties} \\
        \noalign{\smallskip}
        \hline
        2D (Branch Diagram) & 0.8295 & 0.001 & \scinot{4.318}{-6} & 0.1704 \\
        3D (pEWs+$v_{Si~II}$) & 0.6481 & 0.0044 & 0.0015 & 0.3459 \\
        3D (pEWs+$M_B$) & 0.0331 & 0.0014 & \scinot{1.369}{-53} & 0.9655 \\
        4D (pEWs+$v_{Si~II}$+$M_B$) & 0.0203 & 0.0052 & \scinot{1.741}{-9} & 0.9745 \\
        \hline
    \end{tabular}
\end{table}

\end{appendix}

\end{document}